\documentclass[12pt,a4paper]{article}
\pdfoutput=1
\usepackage{epsfig,graphicx,color}
\usepackage{amssymb}
\usepackage{cite}
\usepackage{amsfonts}
\usepackage{amstext}
\usepackage{amsmath}

\newcommand{\beq}{\begin{equation}}
\newcommand{\eeq}{\end{equation}}
\newcommand{\be}{\begin{equation}}
\newcommand{\ee}{\end{equation}}
\newcommand{\ba}{\begin{array}}
\newcommand{\ea}{\end{array}}
\newcommand{\beqa}{\begin{eqnarray}}
\newcommand{\eeqa}{\end{eqnarray}}
\newcommand{\bea}{\begin{eqnarray}}
\newcommand{\eea}{\end{eqnarray}}

\newcommand{\cO}{{\cal O}}
\newcommand{\cB}{{\cal B}}

\newcommand{\cL}{{\cal L}}
\newcommand{\Tr}{{\rm Tr}}
\newcommand{\re}{{\rm Re}}
\newcommand{\im}{{\rm Im}}

\newcommand{\no}{\nonumber}

\newcommand{\YU}{Y_u}
\newcommand{\YD}{Y_d}
\newcommand{\YUD}{Y_{u(d)}}
\newcommand{\PU}{P_u}
\newcommand{\PUD}{P_{u(d)}}
\newcommand{\PD}{P_d}
\newcommand{\Vt}{\widetilde V}
\newcommand{\Vtr}{\widetilde V_0}
\newcommand{\ts}{\tilde s}
\newcommand{\tc}{\tilde c}

\newcommand{\dbase}{~\big|_{d-{\rm base}}}
\definecolor{red}{cmyk}{0,1,1,0.4}
\definecolor{blue}{cmyk}{1,1,0,0}
\newcommand{\Heff}{{\cal H}_\text{eff}}
\newcommand{\gev}{\, {\rm GeV}}
\DeclareMathOperator{\RE}{Re}
\DeclareMathOperator{\IM}{Im}

\begin{document}

\begin{flushright}  
TUM-HEP-765/10
\end{flushright}

\medskip

\begin{center}
{\Large \bf \boldmath Quark flavour mixing with right-handed currents: \\
an effective theory approach}\\[0.8 cm]
{\large Andrzej~J.~Buras$^{a,b}$, Katrin Gemmler$^{a}$, Gino Isidori$^{b,c}$} \\[0.5 cm]
\small
$^a${\em Physik-Department, Technische Universit\"at M\"unchen, James-Franck-Stra{\ss}e,
\\D-85748 Garching, Germany} \\[0.1cm]
$^b${\em TUM Institute for Advanced Study, Technische~Universit\"at~M\"unchen, Arcisstra{\ss}e 21,
\\D-80333 M\"unchen, Germany} \\[0.1cm]
$^c${\em INFN, Laboratori Nazionali di Frascati, I-00044 Frascati, Italy} \\[0.8 cm]
\end{center}

\abstract{%
\noindent 
The impact of right-handed currents in both charged- and neutral-current 
flavour-violating processes is analysed by means of an effective 
theory approach. More explicitly, we analyse the structure of dimension-six
operators assuming a left-right symmetric flavour group, commuting with 
an underlying $SU(2)_L \times SU(2)_R \times U(1)_{B-L}$ global symmetry, 
broken only by two Yukawa couplings. The model contains a new 
unitary matrix controlling flavour-mixing in the right-handed sector. 
We determine the structure of this matrix by charged-current data, 
where the tension between inclusive and exclusive determinations of 
$|V_{ub}|$ can be solved. Having determined the size and the flavour structure of 
right-handed currents, we investigate how they would manifest themselves in 
neutral current processes, including particle-antiparticle mixing, 
$Z\to b \bar b$, $B_{s,d}\to \mu^+\mu^-$, $B\to \{X_s,K,K^*\} \nu\bar\nu$,
and $K\to \pi\nu\bar\nu$
decays. The possibility to explain a non-standard CP-violating phase 
in $B_s$ mixing in this context, and the comparison with other 
predictive new-physics frameworks addressing the same problem, 
is also discussed. While a large $S_{\psi\phi}$ asymmetry can easily 
be accommodated, we point out a tension 
in this framework between $|V_{ub}|$ and $S_{\psi K}$. }


\section{Introduction}

 One of the main properties of the Standard Model (SM) regarding flavour 
 violating processes is the left-handed 
 structure of the charged currents that is 
 in accordance with the maximal violation 
 of parity observed in low energy processes. Left-handed charged currents encode
 at the level of the Lagrangian the full information about flavour mixing 
 and CP violation represented compactly by the CKM matrix.
 Due to the GIM
 mechanism this structure has automatically profound implications for
 the pattern of FCNC processes that seems to be remarkably
 in accordance with the 
 present data within theoretical and experimental uncertainties, bearing in 
 mind certain anomalies which will be discussed below.

 Yet, the SM is expected to be only the low-energy limit of a more fundamental
 theory in which  parity could be a good symmetry implying the existence of
 right-handed charged currents. Prominent examples of such fundamental 
 theories are  left-right symmetric models on which a rich literature exists.

 Left-right symmetric models were born 35 years 
 ago~\cite{Pati:1974vw,Mohapatra:1974gc,Mohapatra:1974hk,Senjanovic:1975rk,Senjanovic:1978ev}
 and extensive
 analyses of many observables can be found in the literature~(see 
 e.g.~\cite{Zhang:2007da,Maiezza:2010ic} and references therein).
 Renewed theoretical interest in models with an underlying $SU(2)_L \times SU(2)_R$ 
 global symmetry has also been motivated by Higgsless 
models~\cite{Csaki:2003zu,Nomura:2003du,Barbieri:2003pr,Georgi:2004iy}. However,
 the recent phenomenological interest in making another look  at the right-handed 
 currents in general, and not necessarily in the context of a given left-right 
 symmetric model, originated in tensions between inclusive and exclusive
 determinations of the elements of the CKM matrix $|V_{ub}|$ and  $|V_{cb}|$. 
 It could be that these tensions are due to the underestimate of theoretical 
 and/or experimental uncertainties. Yet, it is a fact, as pointed out
 and analyzed recently in particular in Ref.~\cite{Crivellin:2009sd,Chen:2008se}, 
 that the presence of right-handed  currents could either remove 
 or significantly weaken some of these tensions, especially in the 
 case of $|V_{ub}|$.

 Assuming that right-handed currents provide the solution to
 the problem at hand, there is an important question whether the strength
 of the right-handed currents required for this purpose is consistent with
 other observables and whether it implies new effects somewhere else that
 could be used to test this idea more globally.

 In the present paper we make still another look at the effects of 
 the right-handed currents in low energy processes, without specifying
 the fundamental theory in detail but only assuming its global symmetry
 and the pattern of it breakdown. Specifically we address the following
 questions:
 \begin{itemize}
 \item 
 What is the allowed structure of the right-handed matrix that 
 governs flavour violating processes in the right-handed sector 
 and reduces the tension observed in the inclusive and exclusive 
 determinations of $|V_{ub}|$?
 \item
 What is the impact of right-handed currents, in combination with
 SM left-handed currents, on particle-antiparticle mixing?
 \item
 Can in this context the known  $Z\to b \bar b$ ``anomaly'' be solved?
 \item
 Can such effects be seen in rare FCNC decays such as
 $B_{s,d}\to \mu^+\mu^-$, $B\to \{X_s,K,K^*\} \nu\bar\nu$ and $K \to \pi \nu\bar\nu$?
 \end{itemize}
As there have been already numerous analyses of right-handed currents in 
the literature, it is mandatory for us to state what is new in our paper:
\begin{itemize}
\item
In the spirit of~\cite{D'Ambrosio:2002ex} we use an effective theory approach to
describe the effects of right-handed currents in flavour violating processes.
In fact our work could be considered to be a simple generalization of the 
usual MFV framework to include right-handed currents. Indeed the main
model-dependent assumption of our analysis is the hypothesis that 
the left-right symmetric flavour group is broken only by two Yukawa couplings. 
\item
 We determine the new unitary matrix $\tilde V$
 that controls flavour-mixing in the right-handed sector by
 using the data on the tree level charged current transitions $u\to d$,
 $u\to s$, $b\to u$ and $b\to c$ and its unitarity.
 Here the novel feature of our analysis, as compared 
 to~\cite{Crivellin:2009sd,Feger:2010qc},
 is the determination of the full right-handed matrix and not only of its
 $b\to c,u$ elements.
\item
 We point out that the elements of this matrix can be further constrained
 through the FCNC processes, that we analyse in detail. Here our  
 minimalistic assumption about the breaking of the flavour symmetry 
 by only two Yukawa couplings plays a key role, and distinguishes our work
 from most of the existing analyses.
\item
 We point out that the explanation of the different values of $|V_{ub}|$ 
 from inclusive and exclusive semi-leptonic decays with the help of
 right-handed currents, implying large value of $|V_{ub}|$, strengthens  
 the tension (already existing in the SM) between $\sin 2\beta$ 
 and $S_{\psi K_S}$. This tension cannot be 
 solved through the new CP-violating effects in the right-handed matrix as
 the contributions of right-handed currents to $B_d^0-\bar B_d^0$ mixing turn
 out to be strongly suppressed due to the $\varepsilon_K$ constraint
 and the 
 desire to explain the enhanced value of the $S_{\psi\phi}$ asymmetry in 
 the $B_s$ system.
\end{itemize}

 Our paper is organized as follows. In Section~2 we present the electroweak 
 symmetry and the  field  content of our 
 effective theory. Here we also introduce the
 right-handed (RH) matrix $\Vt$ and discuss some of its properties. In Section~3
 we make a closer look at the RH charged currents. 
 In Section 4 we use the present
 knowledge on $u\to d$, $s\to d$, $b\to c$ and $b\to u$ transitions to 
 obtain upper bounds on most of the elements of the matrix $\Vt$. 
 Section~5 is devoted to a general discussion of dimension-six operators 
 relevant to neutral currents. The impact of these operators is then 
 analysed in particle-antiparticle mixing (Section 6) and in 
 processes sensitive to $Z$-mediated neutral currents (Section 7). 
 A comparison of the results obtained in the present framework
 with those obtained under the MFV hypothesis, and in explicit 
 left-right models, is presented  in Section 8. 
We summarize our main results and conclude in Section~9.

\section{The model} 

\subsection{Electroweak symmetry and field content}

The starting point of our analysis is the assumption that 
the SM is the low-energy limit of a more
fundamental theory. We don't know the exact structure of this
theory, but we assume that in the high-energy limit it is 
left-right symmetric. The difference of left-handed (LH) and 
 right-handed  sectors observed in the SM is 
only a low-energy property, due to appropriate symmetry-breaking 
terms.

In particular, we assume that the theory has 
a $SU(2)_L \times SU(2)_R \times U(1)_{B-L}$ global 
symmetry, explicitly broken only in the Yukawa 
sector and by the $U(1)_Y$ gauge coupling. 
Under this symmetry the SM quark fields can be grouped 
into three sets of LH or RH doublets with $B-L$ charge $1/3$:
\be
\qquad 
Q^i_{L} = \left( \ba{c} u_{L}^i \\ d_{L}^i \ea \right)~, \qquad 
Q^i_{R} = \left( \ba{c} u_{R}^i \\ d_{R}^i \ea \right)~, \qquad 
i=1 \ldots 3~.
\ee
With this assignment the SM hypercharge is given by $Y=T_{3R} + (B-L)/2$.  
In order to recover the SM electroweak gauge group,  
we assume that only the $SU(2)_L$ and $U(1)_Y$ subgroups 
of $SU(2)_L \times SU(2)_R \times U(1)_{B-L}$ are effectively 
gauged below the TeV scale. In close analogy we can 
introduce three sets of LH and RH 
leptons, $L^i_L$ and $L_R^i$ (including three 
RH neutrinos), with $B-L=-1$.

The breaking of the electroweak symmetry is achieved via 
the spontaneous breaking of $SU(2)_L \times SU(2)_R$ 
into the vectorial subgroup $SU(2)_{L+R}$ at the electroweak 
scale. For simplicity, we provide an explicit description 
of this breaking introducing a SM-like Higgs field 
transforming as ($2,\bar 2$) of $SU(2)_L \times SU(2)_R$, 
with non-vanishing vacuum expectations value (vev):
\be
H \to \mathcal{U}_L H \mathcal{U}_R^\dagger~, \qquad  \mathcal{U}_{L(R)} \in  SU(2)_{L(R)}~, \qquad
\langle H \rangle = v \left(\ba{cc}1 & 0 \\ 0 & 1 \ea\right)~.
\ee
Introducing the kinetic term 
\be
\cL^{\rm kin}_{\rm Higgs} = \frac{1}{4}
\Tr[(D_\mu H)^\dagger D_\mu H]~,
\ee
where $D_\mu H =\partial_{\mu} H -i g W^a_{\mu}T_a H+ig^\prime HT_{3}B_\mu$,
we recover the standard tree-level expressions of $W$ and $Z$ masses 
for $v \approx 246$~GeV. 
However, most of the following discussion on 
flavour-violating effective operators applies as well to models 
where the $SU(2)_L \times SU(2)_R \to SU(2)_{L+R}$ breaking is 
achieved via a more complicated Higgs sector, or even without 
a fundamental Higgs field. 
In the latter case $H$ is replaced by 
$v\times U$, where $U$ is the unitary field, transforming
as ($2,\bar 2$) of $SU(2)_L \times SU(2)_R$, 
that encodes the three Goldstone bosons (see e.g.~Ref.~\cite{Isidori:2009ww}).

\subsection{The quark flavour symmetry}

As far as the quark flavour structure is concerned, 
we assume an $SU(3)_L \times SU(3)_R$ flavour symmetry, 
with the fields transforming as
\be
Q_{L(R)} \to f_{L(R)} Q_{L(R)}~, \qquad f_{L(R)} \in  SU(3)_{L(R)}~.
\label{eq:QR}
\ee
In order to generate different masses for up- and down-type 
quarks we introduce two spurion fields, $\PUD$,
transforming as $\mathcal{U}_R \PUD \mathcal{U}_R^\dagger$ under $SU(2)_{R}$, 
whose background values break the custodial $SU(2)_{L+R}$ 
symmetry:
\be
\PU  = \left(\ba{cc}1 & 0 \\ 0 & 0 \ea\right)~, 
\qquad
\PD  =  \left(\ba{cc}0 & 0 \\ 0 & 1 \ea\right)~.
\ee
In addition, we assume that the flavour symmetry is broken by two spurions,
both transforming as $(3,\bar 3)$ under $SU(3)_L \times SU(3)_R$:
\be
\YUD \to f_L ~\YUD~ f_R^\dagger ~.
\ee
Finally, we assume an additional $U(1)$ symmetry, 
under which $\YU$ and $\YD$ have different charges, 
and $\PU$ and $\PD$ have the corresponding opposite 
charges.
This symmetry structure implies the following
invariant quark Yukawa interactions:
\be
\cL_{\rm Y} =   \bar Q_L H \YU  \PU Q_R  + \bar Q_L H \YD  \PD Q_R  + {\rm h.c.}
\label{eq:Yd4}
\ee
This is equivalent to the SM Lagrangian when we take into account 
the structure of $\PU$ and $\PD$. In models 
with two Higgs doublets the two spurions $\PU$ and $\PD$ can
be interpreted as the remnant of the different vevs 
of the two Higgs fields. Assuming heavy masses
and similar vevs for these Higgs fields, 
this picture cannot be distinguished 
from the one-Higgs doublet case in our 
effective-theory approach.

Rotating $Q_{L}$ and $Q_{R}$ in flavour space we can always choose 
a quark basis where one of the two Yukawa couplings is diagonal.
We can also rotate the relative phases of the quark fields to make 
this diagonal matrix real. Choosing the basis where $\YD$ is diagonal we can
write
\bea
\YD\dbase &=& \lambda_d~, \qquad\qquad     
              \lambda_d = \frac{\sqrt{2}}{v} {\rm diag} (m_{d}, m_{s}, m_{b} ) 
\equiv {\rm diag} (y_{d}, y_{s}, y_{b} ) ~, \no \\
\YU\dbase &=& V^\dagger \lambda_u \Vt,\qquad\quad  
              \lambda_u = \frac{\sqrt{2}}{v} {\rm diag} (m_{u}, m_{c}, m_{t} )
\equiv {\rm diag} (y_{u}, y_{c}, y_{t} ) ~,
\eea
where $V$ and  $\Vt$ are two unitary complex $3\times 3$ mixing matrices. 
$V$ can be identified with the CKM matrix, while $\Vt$ is a new unitary 
mixing matrix that parametrizes 
the misalignment of $\YU$ and $\YD$ in the RH sector.
In such a basis, compatible with the $SU(3)_L \times SU(3)_R$ flavour symmetry, 
the mass terms generated by $\cL_{\rm Y}$ once the Higgs gets a vev
are:
\be
\cL^{\rm mass}_{\rm Y} =   v \bar u_L V^\dagger \lambda_u \Vt u_R  
+ v \bar d_L \lambda_d d_R  + {\rm h.c.}
\ee
In order to diagonalize the mass terms for the up-quarks we need
to perform the following (flavour-breaking) rotations 
\be
 u_L \to  u^\prime_L = V u_L~, \qquad 
 u_R \to  u_R^\prime = \Vt u_R~,
\label{eq:RHrot}
\ee
with $ u^\prime_{L,R}$ denoting the mass-eigenstate fields.

\subsection{RH mixing matrix and FCNC spurions}
The new mixing matrix $\Vt$  can be parametrized in terms 
of 3 real mixing angles
and 6 complex phases. Adopting the standard CKM phase convention, 
where the 5 relative phases of the quark  fields are adjusted to 
remove 5 complex phases from the CKM matrix, we have no more
freedom to remove the 6 complex phases from  $\Vt$.  In the
standard CKM basis  $\Vt$ can be 
parametrized as follows
\be
\Vt = D_U \Vtr D^\dagger_D~, 
\ee
where $\Vtr$ is a ``CKM-like'' mixing matrix, containing only one non-trivial phase
and $D_{U,D}$ are diagonal matrices containing the remaining CP-violating phases.
For reasons that will become clear in the following, it is convenient to 
attribute the non-trivial phase of  $\Vtr$ to the $2-3$ mixing, 
such that 
\be
 \Vtr = 
\left(\begin{array}{ccc}
\tc_{12}\tc_{13}&\ts_{12}\tc_{13}&\ts_{13}\\ -\ts_{12}\tc_{23}
-\tc_{12}\ts_{23}\ts_{13}e^{-i\phi} &\tc_{12}\tc_{23}-\ts_{12}\ts_{23}\ts_{13}e^{-i\phi}& 
\ts_{23}\tc_{13}e^{-i\phi}\\ 
-\tc_{12}\tc_{23}\ts_{13}+\ts_{12}\ts_{23}e^{i\phi}& -\ts_{12}\tc_{23}\ts_{13}-\ts_{23}\tc_{12}e^{i\phi}&\tc_{23}\tc_{13}
\end{array}\right)~,
\label{eq:Vtrgen}
\ee
and 
\be
D_U={\rm diag}(1, e^{i\phi^u_2}, e^{i\phi^u_3})~,  \qquad 
D_D={\rm diag}(e^{i\phi^d_1}, e^{i\phi^d_2}, e^{i\phi^d_3})~.
\label{eq:Dphases}
\ee

Combining the basic spurions $\YU$ and  $\YD$ we can build 
symmetry-breaking terms transforming as $(8,1)$ or $(1,8)$ 
under the $SU(3)_L \times SU(3)_R$ flavour group. These 
spurions control the strength of FCNCs in the model. Since 
the only large Yukawa coupling is the top-one, only two 
of such terms are phenomenologically relevant: 
$\YU \YU^\dagger \sim (8,1)$ and $\YU^\dagger \YU \sim (1,8)$.
Their explicit expressions in the mass-eigenstate basis of
down quarks are
\bea
(\YU \YU^\dagger)_{i\not=j} \dbase  &=& (V^\dagger \lambda_u^2 V)_{ij}
\approx y_t^2 V^*_{3i} V_{3j}~,  \\
(\YU^\dagger \YU)_{i\not=j} \dbase  &=& (\Vt^\dagger \lambda_u^2 \Vt)_{ij}
\approx y_t^2 
e^{i(\phi^d_i-\phi^d_j)}(\Vt_0)^*_{3i} (\Vt_0)_{3j}~.
\label{eq:YYR}
\eea
The $\YU \YU^\dagger$ term, which appears in LH mediated 
FCNCs, has exactly the same structure as in the MFV framework~\cite{D'Ambrosio:2002ex}.
The  $\YU^\dagger \YU$ term is a new spurion characterizing 
the strength of RH mediated FCNCs in this model.
To make contact with the analysis of Ref.~\cite{Feldmann:2006jk}, where 
the MFV flavour group ($ SU(3)_{Q_L} \times SU(3)_{u_R}\times SU(3)_{d_R}$) 
with non-minimal breaking terms has been considered, our framework 
corresponds to the introduction of a single non-MFV spurion, $\Vt$, 
that transforms as $(\bar 3,\bar 3)$ under $SU(3)_{u_R}\times SU(3)_{d_R}$
and it is constrained to be a unitary matrix. 

\section{A first look at the dimension-six operators}

\subsection{Preliminaries} 
As anticipated, we do not specify the ultraviolet completion of the model.
We proceed encoding the effects of the high-energy 
degrees of freedom by means of the effective Lagrangian 
\be
\cL_{\rm eff}  =  \cL_{\rm SM} +  \frac{1}{\Lambda^2} \sum c_i  O^{(6)}_i~,
\ee
where the $O^{(6)}_i$ are dimension-six effective operators compatible with the 
symmetries discussed before. Here $\Lambda$ is an effective scale, expected to 
be of $\cO(1~{\rm TeV})$, and the $c_i$ are dimensionless coefficients. 

In order to build the basis of relevant effective operators, 
it is first convenient to look at the quark bilinear currents 
invariant under the flavour symmetry defined above. 
Introducing terms with at most two Yukawa spurions
(with no more than one $\YD$), and denoting with $\Gamma$ 
a generic Dirac structure, we have
\bea
\cO(Y^0): &\qquad&  \bar Q_L \Gamma Q_L~, \qquad\qquad\ \! 
                    \bar Q_R \Gamma Q_R~, \label{eq:Y0} \\
\cO(Y^1): &\qquad&  \bar Q_L \Gamma \YU \PU Q_R~,  \qquad 
                    \bar Q_L\Gamma \YD \PD Q_R~, \label{eq:Y1} \\
\cO(Y^2): &\qquad&  \bar Q_L  \Gamma \YU\YU^\dagger  Q_L~, \qquad 
                    \bar Q_R \Gamma \YU^\dagger\YU  Q_R~.  \label{eq:Y2}
\eea
Most of these bilinear structures are allowed
also in the MFV case. The only two notable exceptions are: 
i) the charged-current component of the  
RH bilinear $\bar Q_R \Gamma Q_R$, and 
ii) charged-  and neutral-current components of 
$\bar Q_R  \Gamma  \YU^\dagger\YU Q_R$.
These two will play the central role in our paper. We 
first discuss charged currents, postponing the analysis of 
neutral currents to Section~\ref{sect:d6fcnc}.

\subsection{Modification of charged currents}
\label{sect:cc1}
 
Our first goal is to understand which operators can probe the 
rotation in the RH sector, namely the flavour-mixing matrix $\Vt$
that appears in the charged-current component of the  
bilinear $\bar Q_R \Gamma Q_R$. 
If we consider operators with only two quark fields, and we 
ignore RH neutrinos (assuming they are heavy),
the list of relevant operators is quite small:
\bea
&& O^{(6)}_{R_{\ell1}} = \bar Q_R \gamma^\mu \tau_i  Q_R~\bar L_L \gamma_\mu \tau^i  L_L~, \no \\
&& O^{(6)}_{R_{h1}} = i\bar Q_R \gamma^\mu H^\dagger D_\mu H Q_R~, \qquad 
   O^{(6)}_{R_{h2}} = i\bar Q_R \gamma^\mu \tau_i Q_R~\Tr\left( H^\dagger D_\mu H \tau^i \right)~.
\label{eq:RHops1}
\eea
Most important, all these operators 
are equivalent as far as the quark-lepton charged-current 
interactions are concerned. In the case of $O^{(6)}_{R_{hi}}$ we generate an effective 
coupling of the RH quark current to the $W$ field 
after the breaking of the electroweak symmetry:
integrating out the $W$ leads to a quark-lepton charged-current 
interaction identical to the one of $O^{(6)}_{R_{\ell1}}$.

The resulting effective quark-lepton charged-current 
interaction obtained integrating out the $W$ 
at the tree-level can be written as
\be
\cL^{c.c.}_{\rm eff} =  \left( - \frac{g^2}{2 M_W^2} +   \frac{c_{L} }{\Lambda^2} \right)
~\bar u_L \gamma^\mu d_L ~\bar \ell_L \gamma_\mu \nu_L
+ \frac{c_{R} }{\Lambda^2} ~
\bar u_R \gamma^\mu d_R  ~ \bar \ell_L \gamma_\mu \nu_L~+~{\rm h.c.} \,.
\ee
In the limit $c_L=c_R=0$ we recover the usual SM result. 
The term proportional to $c_R$ is the result 
of the new operators in Eq.~(\ref{eq:RHops1}):
$c_R = - 2( c_{R_{h1}} + 2 c_{R_{h2}} -c_{R_{\ell 1}})$.
For completeness, we have also included a possible 
modification of the LH interaction, parametrized by $c_L$.
This is naturally induced by operators obtained from 
Eq.~(\ref{eq:RHops1}) with $Q_R \to Q_L$. 

In principle, charged-current interactions are potentially 
sensitive also to operators written in terms of the bilinears 
in Eqs.~(\ref{eq:Y1})--(\ref{eq:Y2}). However, as long as
we are interested in processes where the up-type quarks are 
of the first two generations, these terms are safely negligible,
being suppressed by small Yukawa couplings.

Rotating the up-type fields to the mass-eigenstate basis 
by means of Eq.~(\ref{eq:RHrot}), and omitting the prime indices for simplicity,
we can finally write
\be
\cL^{c.c.}_{\rm eff} =  - \frac{4 G_F}{\sqrt{2}} ~\bar u \gamma^\mu \left[ (1+\epsilon_L) V P_L  
+ \epsilon_R \Vt P_R \right]  d ~(\bar \ell_L \gamma_\mu \nu_L)~+~{\rm h.c.}
\label{eq:Leffcc}
\ee
where
\be
P_{L} = \frac{1 - \gamma_5}{2}~, \qquad
P_{R} = \frac{1 + \gamma_5}{2}~, 
\ee
\be 
\epsilon_{R} = - \frac{c_{R} v^2 }{2 \Lambda^2} = 
 \frac{v^2}{\Lambda^2}( c_{R_{h1}} + 2 c_{R_{h2}} -c_{R_{\ell 1}})~, \qquad 
\epsilon_{L} = - \frac{c_{L} v^2 }{2 \Lambda^2}~.
\ee

\section{Phenomenology of RH charged currents} 
\label{sect:cc2}

In this section we analyse the phenomenology of RH charged currents.
In particular,  we determine 
the present bounds on the RH mixing matrix
$\Vt$, and  we discuss 
the related impact in the determination of the CKM matrix $V$,
using the effective Lagrangian $\cL^{c.c.}_{\rm eff}$ in Eq.~(\ref{eq:Leffcc}).

Before starting the phenomenological analysis, we recall that 
QED and QCD respects chiral symmetry. As a result, the two operators 
in $\cL^{c.c.}_{\rm eff}$ are not mixed by renormalization group 
effects and are multiplicatively renormalized in the same way.  
This implies that in most cases we can incorporate radiative 
corrections in a straightforward way using SM results. 

\subsection{\boldmath  Bounds from $u\to d$ and $s\to d$ transitions}
Within the SM the leading constraints on $|V_{ud}|$ are 
derived from super-allowed ($0^+\to 0^+$) nuclear beta decays
and by the pion decay ($\pi\to e\nu$) \cite{Amsler:2008zzb}:
\bea
|V_{ud}(0^+ \to 0^+)|^{\rm exp}_{\rm SM} &=& 0.97425 (022)~, \\
|V_{ud}(\pi\to e\nu)|^{\rm exp}_{\rm SM} &=& 0.97410 (260)~.
\eea
By construction, the super-allowed nuclear beta decays are sensitive only 
to the $u\to d$ vector current, while the pion decay is sensitive only  
the $u\to d$ axial current. As a result, the corresponding 
constraints can be implemented in our effective theory via the conditions
\bea
\left| (1+\epsilon_L) V_{ud} + \epsilon_R \Vt_{ud} \right| 
 &=& |V_{ud}(0^+ \to 0^+)|^{\rm exp}_{\rm SM}~, \\
\left| (1+\epsilon_L) V_{ud} - \epsilon_R \Vt_{ud} \right| 
 &=& |V_{ud}(\pi\to e\nu)|^{\rm exp}_{\rm SM}~.
\eea
In principle there is also a constraint from the neutron beta decay.
However, the situation of the neutron life-time measurements is 
quite confusing at present, and this constraint does not add an additional 
significant new information.

Since we expect $\epsilon_{L,R} \ll 1$, we can expand the above equations 
to first order in  $\epsilon_{L,R}$. For simplicity, we also assume $\epsilon_{L,R}$
to be real. Solving the above constraints under these conditions 
leads to 
\bea
\left| (1+\epsilon_L) V_{ud} \right| = 0.9742 \pm 0.0013~,  \qquad
\epsilon_R~ \re\left(\frac{ \Vt_{ud} }{ V_{ud}}\right) = (0.1 \pm 1.3)\times 10^{-3}~.
\label{eq:Vud1}
\eea

The leading constraints on the vector and the axial $s\to u$ currents 
are derived from $K\to \pi\ell\nu$ and  $K\to \mu\nu$ decays, respectively.
Using the SM results obtained in~\cite{Antonelli:2010yf}, 
\bea
|V_{us}(K\to \pi\ell\nu)|^{\rm exp}_{\rm SM} &=& 0.2243 (12)~, \\
|V_{us}(K\to \mu\nu)|^{\rm exp}_{\rm SM} &=& 0.2252 (13)~, 
\eea
we can impose the following conditions
\bea
\left| (1+\epsilon_L) V_{us} + \epsilon_R \Vt_{us} \right| 
 &=& |V_{us}(K\to \pi\ell\nu)|^{\rm exp}_{\rm SM}~, \\
\left| (1+\epsilon_L) V_{us} - \epsilon_R \Vt_{us} \right| 
 &=& |V_{us}(K\to \mu\nu)|^{\rm exp}_{\rm SM}~. 
\eea
Proceeding as in the $u\to d$ case we find
\bea
\left| (1+\epsilon_L) V_{us} \right| = 0.2248 \pm 0.0009~,  \qquad
\epsilon_R~ \re\left(\frac{ \Vt_{us} }{ V_{us}}\right) = -(2.0 \pm 3.9)\times 10^{-3}~.
\label{eq:Vus1}
\eea

Since we know that $|V_{ub}| =\cO(10^{-3})$, we can neglect $|V_{ub}|^2$
in the CKM unitarity relation $|V_{ud}|^2 + |V_{us}|^2 + |V_{ub}|^2 = 1$,
and impose that $|V_{ud}|^2 + |V_{us}|^2 = 1 + \cO(10^{-4})$.
This allows us to obtain a  determination of $\epsilon_L$ at the $10^{-3}$
level starting from the constraints on $| (1+\epsilon_L) V_{ud(s)}|$ 
in Eq.~(\ref{eq:Vud1}) and (\ref{eq:Vus1}):
\be
\epsilon_L  = 
\left[ (1+ \epsilon_L)^2 (  |V_{ud}|^2 + |V_{us}|^2 + |V_{ub}|^2 ) \right]^{1/2}  - 1
= - (0.2 \pm 1.2) \times 10^{-3}~. 
\label{eq:epsL}
\ee
Using this result back in  Eq.~(\ref{eq:Vud1}) and (\ref{eq:Vus1}) 
we finally obtain:
\bea
&& |V_{ud}| = 0.9742 \pm 0.0013~, \qquad 
\epsilon_R~ \re(\Vt_{ud}) =  (0.1 \pm 1.3)\times 10^{-3}~, \\
&& |V_{us}| = 0.2248 \pm 0.0009~, \qquad 
\epsilon_R~ \re(\Vt_{us}) = -(0.5 \pm 0.9)\times 10^{-3}~.
\label{eq:VusVudR}
\eea
Note that if $\epsilon_R = \cO(10^{-3})$, as expected by na\"ive dimensional
analysis for new-physics at the TeV scale, the above results 
do not imply small mixing angles among the first two generations 
in the RH sector. Similar conclusions from the analysis or RH currents 
in semileptonic $K$ decays were reached also in~\cite{Bernard:2007cf}.
As we will show in the following, this expectation is confirmed also 
by our analysis of FCNC processes.

\subsection{\boldmath  Bounds from $b\to c$ and $b\to u$ transitions}
\label{sect:ccbounds}

Given the smallness of $\epsilon_L$ derived in Eq.~(\ref{eq:epsL}),
we can neglect it in the determination of the 
matrix elements entering $b\to c$ and $b\to u$ 
transitions, where the best experimental errors are 
at least of $\cO(1\%)$. 

First we summarize the bounds from $b\to c$ transitions, starting with the consideration of the inclusive decay $B \to X_c \ell \nu_{\ell}$. The total rate is easy to handle as the obtained correction, being dependent 
on the fraction of the RH mixing matrix element and the CKM matrix element, 
can be factorized in comparison to the SM tree level decay width. 
We proceed in the following way:
the value for $V_{cb}$, obtained from the comparison of the SM decay width and the experimental data, must be equivalent to our effective $V_{cb}$ including the RH contributions. We then obtain
\begin{equation}
\big(|V_{cb}|_{\text{SM-exp}}^{\text{incl}}\big)^2=|V_{cb}|^2 \left[ 1+ |\epsilon_R|^2
 \left|\frac{\Vt_{cb}}{V_{cb}}\right|^2 - r_{\rm int} \, \RE\left(\epsilon_R \frac{\Vt_{cb}}{V_{cb}}\right) 
\right]\,,
\end{equation}
where~\cite{Antonelli:2009ws}
\be
|V_{cb}|_{\text{SM-exp}}^{\text{incl}}=(41.54 \pm 0.73) \times 10^{-3}~,
\label{eq:con1}
\ee
and the strength of the interference $r_{\rm int}$ is given by
\begin{equation}
r_{\rm int}=16\frac{m_c}{m_b} \frac{h(\frac{m_c}{m_b})}{f(\frac{m_b}{m_c})}\,,
\end{equation}
with 
$f(x)=1-8x^2+8x^6-x^8-24x^4 \log x$ and
$h(x)=1-3x^2+3x^4+x^6+6 (x^2+x^4) \log x$. Numerically $r_{\rm int}=0.97\times 10^{-3}$, 
thus the impact of the RH current in the inclusive 
decay turns out to be very small. This is consistent with 
Ref.~\cite{Feger:2010qc}, where a detailed analysis of the inclusive differential distributions
has been performed.

Concerning the exclusive decays  $B \to D^{\ast} \ell \nu_{\ell}$ and $B \to D \ell \nu_{\ell}$, 
the consideration of the differential decay rate turns out to be more useful as here data 
has been determined by various experiments. Furthermore the consideration of the 
heavy-quark limit yields an easy description, 
where just one form factor, the Isgur-Wise function, has to be taken into account close to
the zero-recoil point limit. For completeness, we collect the SM results for the differential 
decay rates close to this kinematical point
\begin{equation}
\frac{d\Gamma(\bar{B}\to D^{\ast} \ell \bar{\nu}_{\ell})}{dw}=
\frac{G_F^2}{48\pi^3} |V_{cb}^{\rm SM}|^2 m_{D^{\ast}}^3 (w^2-1)^{1/2} P(w)|{\cal F}(w)|^2 \,,
\end{equation}
\begin{equation}
\frac{d\Gamma(\bar{B}\to D \ell \bar{\nu}_{\ell})}{dw}=
\frac{G_F^2}{48\pi^3}|V_{cb}^{\rm SM}|^2(m_B+m_D)^2m_D^3(w^2-1)^{3/2}
|{\cal G}(w)|^2\,,
\end{equation}
where in the $B$ meson rest frame $w=E_{D^{(\ast})}/m_{D^{(\ast)}}$ 
and the zero-recoil point limit corresponds to $w=1$. Here $P(w)$ denotes the phase space factor
and ${\cal F}(w)$ and ${\cal G}(w)$ are the hadronic form factors.
For $w=1$, when the momentum transfer of the leptons is at its maximum, $P(1)=12 (m_B-m_{D^{\ast}})^2$. 
From a fit of the kinematical distribution around $w=1$ the experiments determine with high accuracy 
the products  ${\cal F}(1) |V_{cb}|$ and  ${\cal G}(1) |V_{cb}|$
as well as the curvature of the form factors, obtaining~\cite{Antonelli:2009ws},
\bea
{\cal F}(1) |V_{cb}|_{\text{SM-exp}}^{B \to D^{\ast}} &=&(35.41 \pm 0.52)\times 10^{-3} ~, \\
{\cal G}(1) |V_{cb}|_{\text{SM-exp}}^{B\to D} &=&  (42.4 \pm 1.6)\times 10^{-3}~.
\eea
In this kinematical limit 
$B \to D^{\ast} \ell \nu_{\ell}$ and $B \to D \ell \nu_{\ell}$ decays involve only axial and vector 
contributions, respectively. As a result, it is easy to include the RH current contribution.
In analogy to the inclusive case, hence our conditions read
\bea
|V_{cb}|_{\text{SM-exp}}^{B \to D^{\ast}}=|V_{cb}-\epsilon_R \Vt_{cb}|~,\label{eq:con2} \\
|V_{cb}|_{\text{SM-exp}}^{B\to D}=|V_{cb}+\epsilon_R \Vt_{cb}|~.
\label{eq:con3}
\eea
In order to implement these constraints we need to specify the values of the 
form factors at $w=1$.  Using the lattice determinations
${\cal G}(1)=1.074 \pm 0.018 \pm 0.0016$~\cite{Okamoto:2005zg}, 
${\cal F}(1)=0.921 \pm 0.013 \pm 0.0020$~\cite{Bernard:2008dn}, 
leads to 
\be
|V_{cb}|_{\text{SM-exp}}^{B \to D^{\ast}}= (39.4 \pm 1.7) \times 10^{-3}~, \qquad 
|V_{cb}|_{\text{SM-exp}}^{B\to D}=  (38.3 \pm 1.2) \times 10^{-3}~.
\label{eq:con4}
\ee
Performing a global fit to $V_{cb}$ and $\epsilon_R \Vt_{cb}$ using the three constraints 
in Eqs.~(\ref{eq:con1}), (\ref{eq:con2}), and (\ref{eq:con3}), we then obtain
\be
|V_{cb}| = (40.7 \pm 0.6 )\times 10^{-3}, \qquad 
\epsilon_R~ \re\left(\frac{ \Vt_{cb} }{ V_{cb}}\right) = (2.5 \pm 2.5)\times 10^{-2}~,
\ee
with a modest correlation ($\rho=0.16$). This finally implies
\be
\epsilon_R~ \re( \Vt_{cb} ) = (1.0 \pm 1.0)\times 10^{-3}~.
\label{eq:VcbR}
\ee
In this case the $\chi^2$ of the fit is not good ($\chi^2/N_{\rm dof}=4.3$), 
as also in the SM, because both of the exclusive values in Eq.~(\ref{eq:con4})
are below the inclusive one.
This result {\em cannot} be explained in terms of RH currents.
As pointed out in Ref.~\cite{Gambino:2010bp}, the inconsistency among the 
different determinations of $V_{cb}$ is likely to be due to
an overestimate of ${\cal G}(1)$ on the Lattice. Lowering the central value to 
${\cal G}(1)=0.86$, as suggested in \cite{Gambino:2010bp}, and keeping the same error,
leads to a much better fit ($\chi^2/N_{\rm dof}=0.9$).
Since the result for $\epsilon_R~ \re( \Vt_{cb} )$ obtained in this case 
is perfectly consistent with the one in (\ref{eq:VcbR}), in the following we will
use Eq.~(\ref{eq:VcbR}) as reference value. 

\medskip 

We now proceed analysing the constraints from $b\to u$ transitions.
As far as the inclusive rate is concerned, the structure can be obtained 
in a straightforward way from the $b\to c$ case replacing $c \rightarrow u$. 
Here the interference term is totally negligible, so we obtain the condition
\begin{equation}
\big(|V_{ub}|_{\text{SM-exp}}^{\text{incl}}\big)^2= \big(|V_{ub}|^2+ |\epsilon_R|^2
 |\Vt_{ub}|^2 \big)\,,
\end{equation}
where~\cite{Antonelli:2009ws}
\be
|V_{ub}|_{\text{SM-exp}}^{\text{incl}}=(4.11 \pm 0.28) \times 10^{-3}~.
\label{eq:Vub1}
\ee
The inclusive determination from $B\to \pi \ell\nu$, where only the
vector current appears, leads to 
\begin{equation}
|V_{ub}|_{\text{SM-exp}}^{B \to \pi}=|V_{ub}+\epsilon_R \Vt_{ub}| =(3.38 \pm 0.36) \times 10^{-3}~,
\label{eq:Vub2}
\end{equation}
where the experimental value is taken from Ref.~\cite{Antonelli:2009ws}.
Finally, a constraint on the $b\to u$ axial current can be obtained from the 
rare leptonic decay $B\to \tau\nu$. Using the theoretical expression 
\be
\label{eq:BR_B_taunu}
{\cal B}(B\to \tau\nu)_{\rm SM} = 
\frac{G_{F}^{2}m_{B}m_{\tau}^{2}}{8\pi}\left(1-\frac{m_{\tau}^{2}}
{m_{B}^{2}}\right)^{2}f_{B}^{2}|V_{ub}^{\rm SM}|^{2}\tau_{B}~,
\ee
the experimental result 
${\cal B}(B\to \tau\nu)^{\rm exp} = (1.73 \pm 0.34)\times 10^{-4}$~\cite{Bona:2009cj},
and $f_B=(192.8\pm9.9)$~MeV~\cite{Laiho:2009eu}, we get 
\begin{equation}
|V_{ub}|_{\text{SM-exp}}^{B \to \tau}=|V_{ub}-\epsilon_R \Vt_{ub}| = (5.14\pm0.57)\times 10^{-3}~.
\label{eq:Vub3}
\end{equation}

\begin{figure}[t]
\begin{center}
\includegraphics[width=.7\textwidth]{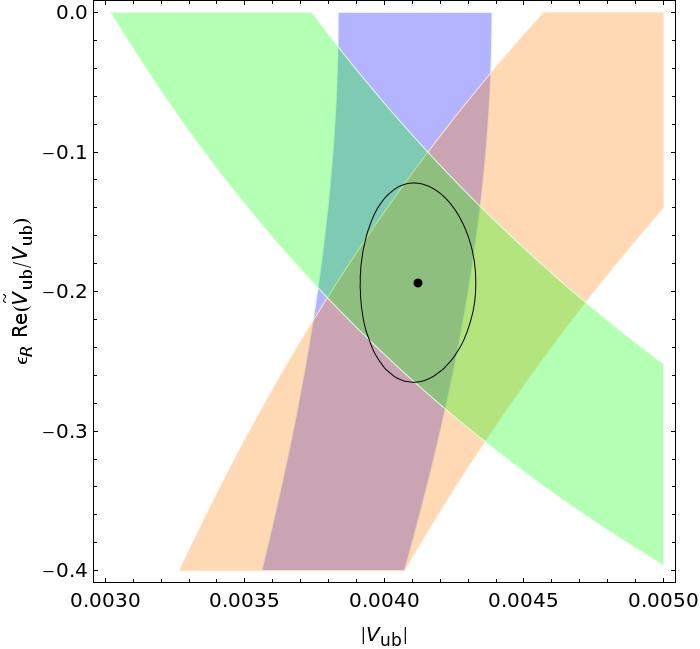}
\vskip -10.8 cm 
{\footnotesize $\qquad\qquad\quad  B\to \pi \ell \nu \qquad\qquad\qquad 
B\to X_u \ell \nu \qquad\qquad B\to \tau\nu$}
\vskip  10.3 cm 
\end{center}
\caption{\label{fig:Vub} Constraints on $|V_{ub}|$ and 
$\epsilon_R~ \re\left(\frac{ \Vt_{ub} }{ V_{ub}}\right)$
from d $B\to \pi \ell \nu$ (green), $B\to X_u \ell \nu$ (blue), and
$B\to \tau\nu$ (orange). 
The bands denote the $\pm 1 \sigma$ intervals of the various experimental
constraints. The ellipse denotes the $1 \sigma$ region of our 
best-fit solution.}
\end{figure}

As noted first in~\cite{Crivellin:2009sd}, here the situation is very favourable for 
the contribution of RH currents, since the axial and vector exclusive determinations are 
substantially above and below the inclusive one (where the interference term is negligible).
Performing a global fit to $V_{ub}$ and $\epsilon_R \Vt_{ub}$ using the 
three constraints we get 
\be
|V_{ub}| = (4.1 \pm 0.2 )\times 10^{-3}, \qquad 
\epsilon_R~ \re\left(\frac{ \Vt_{ub} }{ V_{ub}}\right) = -0.19 \pm 0.07~,
\label{eq:Vubreal}
\ee
with a correlation $\rho=-0.13$, namely an evidence of about 
$2.7\sigma$ of a non-vanishing RH current contribution. 
In this case the quality of the fit is excellent ($\chi^2\approx 0$)
and substantially better than in the absence of RH currents. Most importantly 
the presence of right-handed currents removes the visible discrepancies between 
the various determinations, as shown in Fig.~\ref{fig:Vub}
(the SM case corresponds to the top of the vertical axis, where the 
three determinations of $|V_{ub}|$ are clearly different).

The above result has been obtained expanding to first order in
$\epsilon_R$: in this limit we are sensitive only to the 
combination $\re(\Vt_{ub} /V_{ub})$. 
Given the non-vanishing result for the RH term, we tried also 
a three-parameter fit, where also  $\im( \Vt_{ub}/ V_{ub} )$
is free to vary. In this case no unambiguous result is found, unless additional 
conditions are imposed. Imposing the condition $ |\Vt_{ub}| < |V_{ub}|$,
the best solution is still the one in Eq.~(\ref{eq:Vubreal}), which 
holds for a large interval of $\im(\Vt_{ub} / V_{ub} )$ around zero.
Varying the phase of ${\Vt_{ub}}/{V_{ub}} $ in a conservative 
range leads to 
\be
| \epsilon_R \Vt_{ub} | = (1.0 \pm 0.4)\times 10^{-3}~, 
\qquad {\rm for} \qquad
-\frac{\pi}{4} < {\rm arg}\left(\frac{\Vt_{ub}}{V_{ub}}  \right) < \frac{\pi}{4}~.
\label{eq:VubR}
\ee

\subsection{Global fit of the right-handed mixing matrix}
\label{sect:globalfit}

The information we have collected in the previous section 
can be summarized as follows
\be
| \Vt | \sim  
\left(\begin{array}{ccc}
 < 1.4~ &  < 1.4~ &  1.0\pm 0.4 ~ \\ 
- & - &    < 2.0 ~ \\ - & - &  - 
\end{array}\right) \times \left(\frac{10^{-3}}{\epsilon_R}\right)~,
\label{eq:VRexp}
\ee
where the inequalities correspond to the $\pm 1 \sigma$ interval
and we have assumed small phases except for $\Vt_{ub}$.
The entries without figures have very weak 
direct experimental constraints.
Altogether the constraints in Eq.~(\ref{eq:VRexp}) seem to be rather weak. However, 
thanks the unitarity condition, they are sufficient to draw a series of 
interesting conclusions.
\begin{description}
\item[{\em Constraints on $\epsilon_R$}.] The large value of $|\Vt_{ub}|$ 
allows us to derive a significant constraint on the value of $\epsilon_R$ from 
the unitarity of the first row:
\be
|\epsilon_R|  =   \left( |\epsilon_R \Vt_{ud}|^2 + |\epsilon_R \Vt_{us}|^2 
+ |\epsilon_R \Vt_{ub}|^2  \right)^{1/2} = (1.0 \pm 0.5) \times 10^{-3}~. 
\label{eq:epsR}
\ee
Given the bound on $\epsilon_L$ derived in Eq.~(\ref{eq:epsL}), the 
possibility of $\epsilon_L$ and $\epsilon_R$ of the same order
is perfectly allowed. Note also that the central value in Eq.~(\ref{eq:epsR})
is in good agreement with the na\"ive estimate of models with 
strong electroweak symmetry breaking, where we expect 
$c_{L,R}=\cO(1)$ and $\Lambda=4\pi v \approx 3$~TeV. 

We have no information to disentangle the sign of $\epsilon_{R}$  
and $\Vt_{ub}$. For simplicity in the following we assume $\epsilon_{R}$  
to be positive. This assumption will not have any consequence for
our subsequent analysis since in all observables we 
always have a similar freedom in moving an overall sign from $\Vt$ 
to the Wilson coefficient of the effective operators.

\item[{\em Constraints on $|\Vt_{tb}|$, $|\Vt_{ts}|$, and $|\Vt_{td}|$}.]
Adopting the general parametrization in Eq.~(\ref{eq:Vtrgen}) we 
find a wide range for the three mixing angles compatible with
data. The best-fit solution, obtained using only the constraints 
on $\epsilon_R |\Vt_{ud}|$, $\epsilon_R |\Vt_{us}|$, $\epsilon_R |\Vt_{ub}|$, and
$\epsilon_R |\Vt_{cb}|$, collected in the previous section, is
\be
\Vt  \sim  
\left(\begin{array}{ccc}
0 &  -0.76 &  -0.65\\ 
0.88 & - 0.31& 0.36\\ 
0.48 & 0.57 &  -0.67 
\end{array}\right)~.
\label{eq:matrixfit1}
\ee
However, other solutions with a rather different mixing structure
are also compatible with data. For instance, the hierarchical scenario
with $\ts_{13}\approx 1$ and $\ts_{12}\approx \ts_{23} \approx 0$, 
provides also a good fit. What is interesting is that 
in all cases we can derive non-trivial constraints on the elements 
of the third row, that play a significant role in neutral-current observables
(to be discussed in the next sections). 

From the unitarity condition on the third column, and the large value of $|\Vt_{ub}|$, 
there follows a significant constraint on the maximal value of $|\Vt_{tb}|$. 
A large value of $|\Vt_{tb}|$ is particularly welcome since:
i) it minimizes the values of  $|\Vt_{ts}|$ and $|\Vt_{td}|$, 
that can induce too large contributions to FCNCs; 
ii) it maximizes the impact of right-handed currents in $Z\to b \bar b$,
which could help to improve the agreement with 
experiments (see Sect.~\ref{sect:Zbb}).
In the following we thus concentrate on the scenario of maximal  $|\Vt_{tb}|$. 
This is achieved with the ansatz
\be
 \Vtr^{\rm (I)} = 
\left(\begin{array}{ccc}
\tc_{12}\tc_{13}&\ts_{12}\tc_{13}&\ts_{13} \\ 
-\ts_{12} &\tc_{12} &  0 \\ 
-\tc_{12}\ts_{13} & -\ts_{12}\ts_{13} & \tc_{13}
\end{array}\right)~,
\label{eq:Vtansatz}
\ee
which corresponds to the $\tc_{23}\to 1$ limit 
of the general form in Eq.~(\ref{eq:Vtrgen}).
Using the ansatz (\ref{eq:Vtansatz}), and performing a global fit, 
we find that at the $90\%$~C.L.
\be
|\Vt_{tb}| <  0.73~,
\label{eq:Vubmax}
\ee
where the maximal value 
is obtained for $\ts_{13}\approx-0.68$ and $\ts_{12}\approx-0.84$.
The fact that $|\tilde V_{tb}|$ cannot reach 1 (as in the CKM) implies, in turn,
that $|\Vt_{ts}|$ and $|\Vt_{td}|$ cannot be both vanishing. However, 
the sensitivity of the fit to $\ts_{12}$, which controls their relative strength, 
is very mild. 
The maximal $|\Vt_{tb}|$ in Eq.~(\ref{eq:Vubmax}) is obtained for
$|\Vt_{ts}| \approx |\Vt_{td}| \approx 0.5$, while if we require either 
$|\Vt_{ts}|$ or $|\Vt_{td}|$ to be vanishing, then the maximal 
allowed value of $|\Vt_{tb}|$ decreases by less than $10\%$.
Only large positive values of $\ts_{12}$ are (slightly) disfavored 
by the $\Vt_{us}$ constraint in (\ref{eq:VusVudR}), if $\tc_{13}$ 
is positive.

In summary, we find that the scenario 
with maximal $|\Vt_{tb}|$ is well described by the ansatz
(\ref{eq:Vtansatz}) with 
\be
 |\tc_{13}| \approx - \ts_{13} \approx 0.7~, \qquad \epsilon_R \approx 1 \times10^{-3}~,
\ee
and free $\ts_{12}$, provided that ${\rm sgn}(\tc_{13}\ts_{12})=-1$.
In other words, we find a good description with 
\be
 \Vtr^{\rm (II)} = 
\left(\begin{array}{ccc}
\pm  \tc_{12} \frac{\sqrt{2}}{2} & \pm \ts_{12} \frac{\sqrt{2}}{2} & - \frac{\sqrt{2}}{2} \\ 
-\ts_{12} &\tc_{12} &  0 \\ 
\tc_{12} \frac{\sqrt{2}}{2} & \ts_{12} \frac{\sqrt{2}}{2} &  \pm \frac{\sqrt{2}}{2}
\end{array}\right)~.
\label{eq:Vtansatz2}
\ee
In the following we assume the matrix $\Vtr^{\rm (II)}$ as reference structure
for the analysis of FCNCs. As we will discuss in Section~\ref{sect:DF2section}, this structure is 
a key ingredient to generate a sizable non-standard contribution to $S_{\psi\phi}$.
As a result, after we require a large $S_{\psi\phi}$ (as indicated by recent 
 CDF~\cite{Aaltonen:2007he} 
and D0~\cite{Abazov:2008fj,Abazov:2010hv} results),
most of our conclusions will not depend on the choice of this ansatz. 
It should also be stressed that, contrary to the CKM case, having a zero in $\Vtr$
does not prevent non-vanishing CP-violating effects thanks to the extra phases in
Eq.~(\ref{eq:Dphases}).  

As far as FCNCs are concerned, the only significant alternative to 
the ansatz $\Vtr^{\rm (II)}$ is the possibility of a vanishing small 
$|\Vt_{tb}|$, that we can achieve expanding around $\tc_{13} \ll 1$ 
in (\ref{eq:Vtansatz}):
\be
 \Vtr^{\rm (III)} = 
\left(\begin{array}{ccc}
 \tc_{13}\tc_{12} &\tc_{13} \ts_{12} & - 1 \\ 
-\ts_{12} &\tc_{12} &  0 \\ 
 \tc_{12}  & \ts_{12} & \tc_{13}
\end{array}\right)~.
\label{eq:Vtansatz3}
\ee 
As can be seen, this structure is very efficient in escaping all bounds from 
charged currents but for $b\to u$ transitions, where it has a maximal impact. However,
it also implies naturally small effects in $B$ physics (both in $B_s$ and $B_d$ 
meson-anti-meson mixing, as well as on rare $B$ decays) and potentially large
effects in $K$ physics (which are not allowed by data). 
We thus consider it less interesting with respect to the 
ansatz $\Vtr^{\rm (II)}$ in (\ref{eq:Vtansatz2}).

\end{description}

\subsection{Summary}
We conclude this section with a short summary of the 
results obtained from the analysis of charged currents:
\begin{itemize}
\item{}
RH charged currents can help to reduce the tension between 
inclusive and exclusive determinations in $|V_{ub}|$, as recently pointed out in
\cite{Crivellin:2009sd}. 
However, contrary to~\cite{Crivellin:2009sd} and in agreement with~\cite{Feger:2010qc}, 
we find that RH charged currents
do not have a significant impact in the determination of $|V_{cb}|$.
\item{}
The size of the RH charged-current operators necessary to solve the $|V_{ub}|$
problem points towards an effective new-physics scale $\Lambda \approx 3$~TeV.  
\item{}
Thanks to unitarity, the full structure of the RH mixing matrix is 
quite constrained from a global fit of the available constraints. 
In particular, $|\Vt_{tb}|$ is constrained strongly 
by $|\Vt_{ub}|$ and unitarity, which in turn implies non-negligible contributions 
to FCNCs and meson-antimeson mixing through $|\Vt_{ts}|$ and  $|\Vt_{td}|$. 
\item{}
The two representative structures of the RH mixing matrix naturally 
emerge in view of the analysis of FCNCs: the structure
$\Vtr^{\rm (II)}$ in~(\ref{eq:Vtansatz2}) is particularly interesting,
since it could allow large effects in the $B_s$ system, without
being automatically excluded by the tight constraints from 
the $K$ and $B_d$ systems; the structure $\Vtr^{\rm (III)}$
in~(\ref{eq:Vtansatz3}) would imply sizable effects only in the kaon system.
\end{itemize}

%
%
%

\section{Dimension-six operators beyond charged currents}
\label{sect:d6fcnc}

Having constrained the structure of the new mixing matrix from charged 
current processes, we are now ready to analyse the effects generated
within our effective theory in neutral-current processes. 
In particular 
in the next two sections we analyse
the effects in the following set of theoretically clean 
observables: down-type particle-antiparticle mixing
($\varepsilon_K$ and $B_{d,s}$--$\bar B_{d,s}$ mixing),
rare  FCNC decays of $B$ and $K$ mesons with a lepton pair in the final state,
and $Z\to b \bar b$.

The dimension-six effective operators contributing to these processes 
can be constructed by appropriate combinations of the bilinear 
structures in Eqs.~(\ref{eq:Y0})--(\ref{eq:Y2}). 
Rather than presenting lengthy expressions 
containing all possible operators, we limit ourself to analyse 
the impact of the most representative ones. In particular, 
we focus our attention on $\Delta F=2$ operators built in terms of 
the $\bar Q_R \YU^\dagger \YU \gamma^\mu Q_R$ bilinear, and $\Delta F=1$ operators
generating an effective right-handed flavour non-universal 
couplings of the $Z$ boson to down-type quarks. 

\subsection{\boldmath  $\Delta F=2$ processes}

The complete set of gauge-invariant dimension-six operators 
contributing to down-type $\Delta F=2$ amplitudes, with 
the minimum number of Yukawa spurions, is  
\bea
O^{(6)}_{LL} &=&  [\bar Q^i_L (\YU \YU^\dagger )_{ij} \gamma_\mu Q^j_L]^2~, \\ 
O^{(6)}_{RR} &=&  [\bar Q^i_R (\YU^\dagger \YU)_{ij} \gamma_\mu Q^j_R]^2~, \\
O^{(6)}_{LR} &=&  [\bar Q^i_L (\YU \YU^\dagger)_{ij} \gamma^\mu  Q^j_L]
[\bar Q^i_R (\YU^\dagger \YU)_{ij} \gamma_\mu Q^j_R]~.
\eea
The first operator, which is present both in the 
general MFV framework~\cite{D'Ambrosio:2002ex}
and in its constrained version~\cite{Buras:2000dm}
has been widely analysed in the literature. This operator generates
short-distance corrections which have exactly the helicity structure 
and CKM factors of the SM short-distance terms.
As a result, it cannot modify the SM predictions
for the time-dependent CP asymmetries in $B_{d}$ and $B_{s}$ decays~\cite{Buras:2000dm},
in particular $S_{\psi K}$ and  $S_{\psi\phi}$, respectively.

A much richer phenomenology is expected from 
$O^{(6)}_{RR}$ and $O^{(6)}_{LR}$, which are not present in the MFV framework
and through which the new RH mixing matrix enters the game.
For this reason, in the following we focus our attention only on 
these two operators, considering the following $\Delta F=2$ effective 
Lagrangian:
\be
\cL^{\Delta F=2} = \frac{c_{RR}}{\Lambda^2} O^{(6)}_{RR} +
\frac{c_{LR}}{\Lambda^2} O^{(6)}_{LR}~.
\label{eq:DF2effRH}
\ee
Here $c_{RR}$ and $c_{LR}$ are flavour-blind dimension-less coefficients, whose 
size will be discussed below.

\subsection{\boldmath $\Delta F=1$ processes}
The list of operators relevant to $\Delta F=1$ FCNC processes 
with a lepton pair in the final state can be divided into three
categories: 1) operators with two quarks and two Higgs fields;
2) operators with two quarks and two lepton fields;
3) dipole-type operators with two quarks and one SM gauge field.

In the first class we have the following two operators 
\bea
 O^{(6)}_{R_{Z1}} &=& i\bar Q^i_R (\YU^\dagger \YU)_{ij}
\gamma^\mu H^\dagger D_\mu H Q^j_R~,  \no \\
   O^{(6)}_{R_{Z2}} &=& i\bar Q^i_R (\YU^\dagger \YU)_{ij}
 \gamma^\mu \tau_i Q^j_R~\Tr\left( H^\dagger D_\mu H \tau^i \right)~,
\label{eq:Zops}
\eea
and the corresponding LH operators obtained from  (\ref{eq:Zops}) 
through $Q_R \to Q_L$ and $Y_u \leftrightarrow Y_u^\dagger$. 
The latter are of MFV type and have already been analysed in the literature.
On the other hand, the two operators in (\ref{eq:Zops})
give rise to an effective RH coupling of the $Z$ boson of the 
type $\bar d_R^i \gamma^\mu d_R^j Z_\mu$ which is not present 
in the MFV framework. This coupling is particularly
interesting since it allows us to establish connections 
between RH effects in $Z\to b \bar b$ and in rare $K$ and $B$ decays. 
In the following we will analyse these connections by means 
of the effective Lagrangian
\be
\cL^{\Delta F=1} = \frac{c_{R_{Z1}}}{\Lambda^2} O^{(6)}_{R_{Z1}} +
\frac{c_{R_{Z2}}}{\Lambda^2} O^{(6)}_{R_{Z2}}~.
\label{eq:LZgen}
\ee

As far as the other two classes of $\Delta F=1$ operators are concerned:
the operators with two quarks and two lepton fields do not lead to 
effects qualitatively different that those obtained after integrating 
out the $Z$ in Eq.~(\ref{eq:Zops}). The left-right operators with 
the photon field are tightly constrained by $B\to X_s\gamma$ and,
similarly to the MFV case, once the $B\to X_s\gamma$ constraint is 
imposed they do not lead to significant effects in other processes.

\section{Meson anti-meson mixing}
\label{sect:DF2section}

\subsection{Preliminaries}
The presence of the right-handed currents in addition to the left-handed ones present in the SM can have considerable impact on particle-antiparticle mixing dominantly through the generation of the LR operators that renormalize strongly under QCD and in the case of $K^0$--$\bar K^0$ mixing have chirally enhanced hadronic matrix elements. Below we present the general structure of the mixing amplitude $M_{12}$ from which the observables like $\varepsilon_K$ and the CP-asymmetries 
$S_{\psi K_S}$ and $S_{\psi \phi}$ can be derived. Similarly $\Delta M_d$ and $\Delta M_s$ can be calculated.

\subsection{\boldmath Effective Hamiltonian for  $\Delta S=2$ transitions}
We will illustrate the full procedure on the example of $\Delta S=2$ transitions. The complete set of operators of dimensions six  involved in the 
presence of left-handed and right-handed currents consists of the following
operators~\cite{Buras:2001ra}:
\begin{eqnarray}
\mathcal{Q}_1^{VLL}&=&\left(\bar s_L\gamma_\mu d_L\right)\left(\bar s_L\gamma^\mu
d_L\right)\,,\nonumber\\
\mathcal{Q}_1^{VRR}&=&\left(\bar s_R\gamma_\mu d_R\right)\left(\bar s_R \gamma^\mu
d_R \right)\,,\nonumber\\
\mathcal{Q}_1^{LR}&=&\left(\bar s_L\gamma_\mu d_L\right)\left(\bar s_R \gamma^\mu
d_R \right)\,,\nonumber\\
\mathcal{Q}_2^{LR}&=&\left(\bar s_R d_L\right)\left(\bar s_L d_R\right)\,,
\label{eq:1.1}
\end{eqnarray}
where we suppressed colour indices as they are summed up in each 
factor $P_{L,R}$.

In the SM only $\mathcal{Q}_1^{VLL}$ contributes. However, in the presence of right-handed currents three additional operators have to be considered. $\mathcal{Q}_1^{VRR}$ renormalizes under QCD, similarly to $\mathcal{Q}_1^{VLL}$, fully independently of other operators and in fact the renormalization group running for the Wilson coefficient of  $\mathcal{Q}_1^{VRR}$ is identical to the one of $\mathcal{Q}_1^{VLL}$: QCD does not  care about the sign of $\gamma_5$. On the other hand $\mathcal{Q}_1^{LR}$ and $\mathcal{Q}_2^{LR}$ mix under renormalization.

Let us next assume that at some scale $\cO$(1~TeV), to be denoted by $\mu_R$, 
the new physics (NP) is integrated out. In the absence of QCD corrections the 
effective Hamiltonian for $\Delta S=2$ corresponding to $\mu_R=\cO(\Lambda)$  
and including only NP contributions reads 
\bea
\left[\Heff^{\Delta S=2}(\mu_R)\right]_{\rm NP}&=&
\frac{1}{\Lambda^2}\big[{C_1}^{VLL}(\mu_R,K)\, 
\mathcal{Q}_1^{VLL}+C_1^{VRR}(\mu_R,K)\,\mathcal{Q}_1^{VRR} 
\no \\ && \qquad
+C_1^{LR}(\mu_R,K)\,\mathcal{Q}_1^{LR} \big]~. 
\label{eq:1.2}
\eea
The following comments should be made:
\begin{itemize}
\item
The coefficients $C_i^a$ with $i=1,2$ and $a=VLL,VRR,LR$ depend generally 
on the system considered ($K$, $B_{s,d}$) as is the case also of the SM.
\item 
At the scale $\mu_R$, before integrating out the $W^{\pm}_L$ fields, 
the Wilson coefficient ${C_1}^{VLL}$ receives only contributions from the NP.
Running down to low scales, ${C_1}^{VLL}$ encodes both SM and NP contributions,
as discussed below.   
\item 
The coefficient $C_2^{LR}(\mu_R,K)$ vanishes at this high scale in the absence of QCD effects 
and it is also vanishing at this scale in the LO renormalization 
group (RG) analysis. However, at NLO it is generally $\cO(\alpha_s(\mu_R))$.
\end{itemize}
Next usually (\ref{eq:1.2}) is evolved by RG down to low energy scales. In this process $C_2^{LR}$ becomes non-vanishing even in the LO approximation. As $\mathcal{Q}_1^{VRR}$ and the complex $(\mathcal{Q}_1^{LR},\mathcal{Q}_2^{LR})$ do not mix with $\mathcal{Q}_1^{VLL}$ and the RG evolution of the latter operator can 
be split into SM and NP part, the NP contribution to the low energy 
effective Hamiltonian at a low scale $\mu_K$ can be immediately evaluated 
by means of analytic expressions in \cite{Buras:2001ra}.

While working with Wilson coefficients and operator matrix elements at
low energy scales is a common procedure, it turns out that   
 for phenomenology it 
is more useful to work directly with $C_i(\mu_R,K)$ and with the 
hadronic matrix elements of  the corresponding operators also evaluated 
at this high scale. The latter matrix elements are given by \cite{Buras:2001ra}
\be\label{eq:matrix}
\langle \bar K^0|Q_i^a|K^0\rangle = \frac{2}{3}M_K^2 F_K^2 P_i^a(K),
\ee
where the coefficients $P_i^a(K)$ 
collect compactly all RG effects from scales below $\mu_R$ as well as
hadronic matrix elements obtained by lattice methods at low energy scales.

With this information at hand we can present compactly basic formulae for 
$\Delta S=2$ and $\Delta B=2$ observables of interest.

\subsection{\boldmath Basic formulae for  $\Delta S=2$ observables}
\label{sect:epsK}

The off-diagonal element in $K^0-\bar K^0$ mixing $M_{12}^K$ can then be decomposed into SM and NP parts
(RH current contributions in our case) 
\begin{equation}
M_{12}^K=\left(M_{12}^K\right)_{\rm SM}+\left(M_{12}^K\right)_{\rm NP},
\end{equation}
where the SM contribution can be found, for instance, in Ref.~\cite{Buras:2001ra},  
and the NP part is obtained from
\begin{equation}\label{eq:M12Heff}
2 m_K \left(M_{12}^K \right)_{\rm NP}^{\ast}=
\langle \bar K^0 | \left[\Heff^{\Delta S=2}(\mu_R)\right]_{\rm NP}|K^0 \rangle \,.
\end{equation}

Using (\ref{eq:1.2}), (\ref{eq:matrix}) and (\ref{eq:M12Heff}) we find
\begin{eqnarray}
\left(M_{12}^K\right)_\text{NP}=\frac{1}{3\Lambda^2}m_K F_K^2\cdot
\Big[\left(C_1^{VLL}(\mu_R,K)+C_1^{VRR}(\mu_R,K)\right)P_1^{VLL}(K)\nonumber\\
+C_1^{LR}(\mu_R,K)P_1^{LR}(K)\Big]^\ast\,,
\label{eq:3.30}
\end{eqnarray}
with
\be
P_1^{VLL}(K) \approx 0.50, \qquad  P_1^{LR}(K) \approx -52 
\ee
obtained by means of analytic formulae in \cite{Buras:2001ra}, with the 
hadronic matrix elements  from~\cite{Babich:2006bh}, the 
updated numerical inputs in Table~\ref{tab:num}, 
and the matching scale $\mu_R=1.5$~TeV.

The $K_L-K_S$ mass difference and the CP-violating parameter $\varepsilon_K$ are given respectively
\begin{eqnarray}
\Delta M_K&=&2\RE M_{12}^K+\left(\Delta M_K\right)_\text{LD} \,,\nonumber\\
\varepsilon_K&=&\frac{\kappa_\varepsilon e^{i\varphi_\varepsilon}}{\sqrt{2}(\Delta M_K)_\text{exp}}\IM M_{12}^K \,,
\end{eqnarray}
where $\varphi_\varepsilon = (43.51\pm0.05)^\circ$  takes into account that $\varphi_\varepsilon\ne \pi/4$ and $\kappa_\varepsilon=0.94\pm0.02$~\cite{Buras:2008nn,Buras:2010pz} includes an additional effect 
from long-distance (LD) contributions (for a recent detailed analysis of $\varepsilon_K$ within the SM see~\cite{Brod:2010mj}).

\begin{table}[t] 
\begin{center}
\begin{tabular}{|l|l||l|l|}
\hline
parameter & value & parameter & value \\
\hline\hline
$F_K$ & $(155.8\pm 1.7)~\text{MeV}$	\cite{Laiho:2009eu}	  & $\Delta M_K$&$(5.292\pm 0.009)
\times 10^{-3} {\rm ps}^{-1}$~\cite{Amsler:2008zzb}\\
$F_{B_d}$ & $(192.8 \pm 9.9)~\text{MeV}$ \cite{Laiho:2009eu} 	  & $\Delta M_d$ & $(0.507\pm 0.005)~ {\rm ps}^{-1}$ ~\cite{Barberio:2008fa}\\
$F_{B_s}$ & $(238.8 \pm 9.5)~\text{MeV}$\cite{Laiho:2009eu}	  & $\Delta M_s$ & $(17.77\pm 0.12)~{\rm ps}^{-1}$ ~\cite{Barberio:2008fa}\\
$\hat B_K$ & $0.725 \pm 0.026$ 	\cite{Laiho:2009eu}		  & $|V_{tb}|$&$ 1\pm 0.06$~\cite{Bona:2007vi}\\
$\hat B_{B_d}$ & $1.26\pm 0.11$ \cite{Laiho:2009eu}		  & $|V_{td}|$&$(8.3\pm 0.5)\times 10^{-3}$~\cite{Bona:2007vi}\\
$\hat B_{B_s}$ & $1.33\pm 0.06$	\cite{Laiho:2009eu}	          & $|V_{ts}|$&$0.040\pm 0.003$~\cite{Bona:2007vi}\\
$M_K$&$0.497614\,\text{GeV}$~\cite{Amsler:2008zzb}		  & $\sin(2\beta_s)$&$0.038\pm 0.003$~\cite{Bona:2007vi}\\ 
$M_{B_d}$&$5.2795$ GeV~\cite{Amsler:2008zzb}                       & $\gamma$  & $1.09\pm 0.12$~\cite{Bona:2007vi}\\
$M_{B_s}$&$5.3664$ GeV~\cite{Amsler:2008zzb}                       & $\varepsilon_K^{\text{exp}}$&$(2.229\pm 0.01)\times 10^{-3}$~\cite{Amsler:2008zzb}\\
$m_t(m_t)$ & $(163.5\pm 1.7) \gev$\cite{:2009ec}		  & $S_{\psi K_S}^{\text{exp}}$& $0.672\pm0.023$~\cite{Barberio:2008fa}\\
\hline
\end{tabular}
\caption{Values of the input parameters used in our analysis of $\Delta F=2$ 
processes.  } \label{tab:num}
\end{center}
\end{table}

Matching the RH effective Lagrangian defined in (\ref{eq:DF2effRH}) 
to the general effective $\Delta S=2$ Hamiltonian in (\ref{eq:1.2}), 
the Wilson coefficients at the high scale read
\begin{eqnarray}
\mathcal{C}_1^{VRR}(\mu_R,K)&=&-c_{RR} y_t^4 e^{2i(\phi^d_2-\phi^d_1)} 
\left[ (\Vt_0)^*_{ts} (\Vt_0)_{td} \right]^2 
\approx -\frac{c_{RR}}{4} y_t^4 e^{2i(\phi^d_2-\phi^d_1)} (\tc_{12} \ts_{12})^2 
\,,\nonumber\\
\mathcal{C}_1^{LR}(\mu_R,K)&=&-c_{LR} y_t^4  e^{i(\phi^d_2-\phi^d_1)} V^*_{ts} V_{td}(\Vt_0)^*_{ts} (\Vt_0)_{td} \approx -\frac{c_{LR}}{2} y_t^4 e^{i(\phi^d_2-\phi^d_1)}
(V^*_{ts} V_{td}) (\tc_{12} \ts_{12})
\,,\nonumber\\
\mathcal{C}_1^{VLL}(\mu_R,K)&=&0\,,
\label{eq:78}
\end{eqnarray}
where the terms on the right-hand side are obtained employing the 
structure (\ref{eq:Vtansatz2}) for the RH mixing matrix.

\begin{table}[t] 
\begin{center}
\begin{tabular}{|c|c||c|c|c|}
\hline
\raisebox{0pt}[10pt][5pt]{Mixing term} & Matrix & $s\to d$  & $b\to d$ & $b\to s$ \\
\hline\hline 
\raisebox{0pt}[12pt][5pt]{   $V^*_{ti} V_{tj}$     } & CKM
&  $ V_{ts}^*V_{td} \approx -\lambda^5 e^{-i\beta} $ 
&  $ V_{tb}^*V_{td} \approx \lambda^3 e^{-i\beta} $ 
&  $ V_{tb}^*V_{ts} \approx -\lambda^2 e^{-i\beta_s} $ \\ \hline\hline
\raisebox{-10pt}[15pt][10pt]{  $\Vt^*_{ti} \Vt_{tj}$  } & 
 $\Vtr^{\rm (II)}$
&  $ \frac{1}{2} \tc_{12} \ts_{12} e^{i(\phi^d_2-\phi^d_1)} $ 
&  $ \pm \frac{1}{2} \tc_{12} e^{i(\phi^d_3-\phi^d_1)} $ 
&  $ \pm \frac{1}{2} \ts_{12} e^{i(\phi^d_3-\phi^d_2)} $ \\  
\raisebox{0pt}[10pt][10pt]{   }                 & 
 $\Vtr^{\rm (III)}$
&  $ \tc_{12} \ts_{12} e^{i(\phi^d_2-\phi^d_1)} $ 
&  $ \tc_{12}\tc_{13} e^{i(\phi^d_3-\phi^d_1)} $ 
&  $ \ts_{12}\tc_{13} e^{i(\phi^d_3-\phi^d_2)} $ \\ 
\hline\hline
\end{tabular}
\caption{Mixing structures relevant to the three down-type $\Delta F=2$ 
and FCNC amplitudes in the SM (LH sector) and in the RH sector.
In the SM case approximate expressions of the CKM factors 
expanded in powers of  $\lambda = |V_{us}|$ are also shown.
In the RH case the two parametrizations for the RH matrix
are those in Eq.~(\ref{eq:Vtansatz2}) and (\ref{eq:Vtansatz3}),
respectively.} \label{tab:YY}
\end{center}
\end{table}

The non-standard contributions to $\Delta S=2$ amplitudes 
are exceedingly large compared to the SM term (and compared 
with data) unless the Wilson coefficients $c_{RR}$ and $c_{LR}$ 
or one of the two mixing terms $\tc_{12}$ or $\ts_{12}$ are very 
small. By construction,  $c_{RR}$ and $c_{LR}$ are flavour-blind and 
therefore the same in the $B_d$ and $B_s$ system. On the other hand, 
the $\tc_{12}$ and $\ts_{12}$ dependencies in the three systems considered are
non-universal, as seen in Table~\ref{tab:YY}, with the observables in the 
$K$-mixing, $B_d$ mixing and $B_s$-mixing dominated by 
 $\tc_{12}\ts_{12}$,  $\tc_{12}$ and $\ts_{12}$, respectively. 
Since both $\Delta S=2$ and $B_d$ mixing are strongly 
constrained, and the data from CDF and D0 give some hints 
for sizable NP contributions in the $B_s$ mixing, 
it is natural to assume in both scenarios for $\Vt$ that $\tc_{12} \ll 1$. 
In this limit the non-vanishing 
Wilson coefficients relevant for $K^0$--$\bar K^0$ mixing and
$\varepsilon_K$ at the high scale read with (\ref{eq:Vtansatz2}) 
\begin{eqnarray}
\mathcal{C}_1^{VRR}(\mu_R,K) &\approx& -\frac{c_{RR}}{4} y_t^4 
e^{2i \phi^d_{21}} \tc_{12}^2 
\,,\nonumber\\
\mathcal{C}_1^{LR}(\mu_R,K) & \approx & -\frac{c_{LR}}{2}  y_t^4
e^{i \phi^d_{21}} V^*_{ts} V_{td}  \tc_{12}~,
\end{eqnarray}
where  $\phi^d_{21}=(\phi^d_2-\phi^d_1)$.
Using these expressions in the above formulae 
for $\Delta M_K$ and $\varepsilon_K$, and taking into 
account the numerical inputs in Table~\ref{tab:num},
we obtain:
\bea
(\Delta M_K)_{\rm RH} & = & (\Delta M_K)_{\rm exp}
\times 
\left[ -2.5\times10^{4} \times  c_{RR} \tc_{12}^2 \cos(2\phi^d_{21})  
\right. \no \\ && \qquad  \left. 
- 1.7\times10^{3} \times c_{LR} \tc_{12}\cos(\phi^d_{21} -\beta+\beta_s) 
\right] \frac{(3~{\rm TeV})^2}{\Lambda^2}~, \\
 (\varepsilon_K)_{\rm RH}
 &=&
 |\varepsilon_K|_{\rm exp} ~e^{i\phi_\epsilon}
\times 
\left[ 3.7 \times10^6 \times c_{RR} \tc_{12}^2 \sin(2\phi^d_{21})  
\right. \no \\ && \qquad  \left. 
+ 2.5\times10^{5} \times  c_{LR} \tc_{12} \sin(\phi^d_{21} -\beta+\beta_s)
\right]
\frac{(3~{\rm TeV})^2}{\Lambda^2}~,
\eea
where, as usual, $\beta$ and $\beta_s$ denote the phases of $V_{td}$ 
and $V_{ts}$ in the standard CKM convention:
\be
V_{td}=|V_{td}|e^{-i\beta}\quad\textrm{and}\quad V_{ts}=-|V_{ts}|e^{-i\beta_s}\,.
\label{eq:3.40}
\ee
As can be seen, the RH contribution is potentially very large and,
independently of the possible value of the CP-violating phase $\phi^d_{21}$,
we get strong constraints on the combinations $c_{RR} \tc^2_{12}$ and 
$c_{LR} \tc_{12}$. In particular, even assuming a vanishing $\phi^d_{21}$,
we get 
\bea
 c_{RR}~\tc_{12}^2 &<& 2.0 \times 10^{-5}~, \qquad {\rm from} \qquad (\Delta M_K)_{\rm RH} 
~<~ 0.5 (\Delta M_K)_{\rm exp}~,
\label{eq:epsbound1} \\
 c_{LR}~\tc_{12} &<& 1.0 \times 10^{-6}~,   \qquad {\rm from} \qquad |\varepsilon_K|_{\rm RH} ~<~ 0.1
|\varepsilon_K|_{\rm SM}~.
\label{eq:epsbound2}
\eea
While the bound on $c_{RR} \tc_{12}^2$ can only become stronger for non-vanishing values 
of $\phi^d_{21}$, the constraint on $c_{LR} \tc_{12}$ can be relaxed to 
$3\times 10^{-4}$ in the fine-tuned scenario where $\phi^d_{21}$ cancels exactly 
the CKM phase of $V^*_{ts} V_{td}$. As we will show in the next section, 
these constraints imply negligible contribution of RH currents to $B_d$ 
mixing, which is one of the important results of our paper.

Before analysing $\Delta B=2$ observables, we briefly discuss what 
happens if do not employ the ansatz (\ref{eq:Vtansatz2}) for the RH mixing 
matrix and, in particular, if we do not assume $\tc_{12} \ll 1$.
As shown in Table~\ref{tab:YY}, employing the structure (\ref{eq:Vtansatz3})
for the RH matrix the mixing structures relevant 
to the kaon system change only by a factor of two. 
As a result, the corresponding bounds on the mixing terms 
are obtained from (\ref{eq:epsbound1})--(\ref{eq:epsbound2})
with the replacement $\tc_{12} \to 2 \tc_{12}\ts_{12}$. 
In this case we can escape the kaon bounds and have sizable 
effects in  $B_d$ mixing if $\tc_{13}$ is not too small.
However, the key ingredient for sizable effects in  $B_d$ mixing
is $\ts_{12} \ll 1$, a configuration that necessarily imply 
small effects in $B_s$ mixing.

\subsection{\boldmath Basic formulae for  $\Delta B=2$ observables}
Similarly to the kaon system, for $B_d^0-\bar B_d^0$ and $B_s^0-\bar B_s^0$ mixing
we can write
\begin{equation}
M_{12}^q=\left(M_{12}^q\right)_{\rm SM}+\left(M_{12}^q\right)_{\rm NP}
\end{equation}
where $q=d,s$. 
For the NP contribution we find 
\begin{eqnarray}
\left(M_{12}^q\right)_\text{NP}=\frac{1}{3\Lambda^2}m_{B_q}F_{B_q}^2
\Big[\left(C_1^{VLL}(\mu_R,B)+C_1^{VRR}(\mu_R,B)\right)P_1^{VLL}(B)\nonumber\\
+C_1^{LR}(\mu_R,B)P_1^{LR}(B)\Big]^\ast\,,
\label{eq:3.31}
\end{eqnarray}
with
\be
P_1^{VLL}(B) \approx 0.70, \qquad  P_1^{LR}(B) \approx -3.2
\ee
obtained by means of analytic formulae in \cite{Buras:2001ra}, with the 
hadronic matrix elements  from~\cite{Babich:2006bh}, the 
updated numerical inputs in Table~\ref{tab:num}, 
and the matching scale $\mu_R=1.5$~TeV.

For the mass differences in the $B_{d,s}^0-\bar B_{d,s}^0$ systems we have
\begin{eqnarray}
\Delta M_q&=&2\left|M_{12}^q\right| \,,\nonumber\\
M_{12}^q&=&\left(M_{12}^q\right)_\text{SM} C_{B_q}e^{2i\varphi_{B_q}}\,,
\qquad (q=d,s)
\label{eq:3.36}
\end{eqnarray}
where
\begin{equation}
\left(M_{12}^d\right)_\text{SM}=\left|\left(M_{12}^d\right)_\text{SM}\right|e^{2i\beta}\,,
\label{eq:3.38}
\end{equation}
\begin{equation}
\left(M_{12}^s\right)_\text{SM}=\left|\left(M_{12}^s\right)_\text{SM}\right|e^{2i\beta_s}\,,
\label{eq:3.39}
\end{equation}
and 
$C_{B_q}\not=1$ and $\varphi_{B_q}\not=0$ summarize the NP effects.

We find then
\begin{equation}
\Delta M_q=(\Delta M_q)_\text{SM}C_{B_q}
\label{eq:3.41}
\end{equation}
and
\begin{equation}
S_{\psi K_S} = \sin(2\beta+2\varphi_{B_d})\,, \qquad
S_{\psi\phi} =  \sin(2|\beta_s|-2\varphi_{B_s})\,,
\label{eq:3.43}
\end{equation}
with the latter two observables being the coefficients of $\sin(\Delta M_d t)$ and $\sin(\Delta M_s t)$ in the time dependent asymmetries in $B_d^0\to\psi K_S$ and $B_s^0\to\psi\phi$, respectively. Thus in the presence of non-vanishing $\varphi_{B_d}$ and $\varphi_{B_s}$ these two asymmetries do not measure $\beta$ and $\beta_s$ but $(\beta+\varphi_{B_d})$ and $(|\beta_s|-\varphi_{B_s})$, respectively.

The non-vanishing Wilson coefficients at the high scale for 
the $B_{s,d}$ systems are
\begin{eqnarray}
\mathcal{C}_1^{VRR}(\mu_R,B_q)&=&-c_{RR} y_t^4 e^{2i(\phi^d_3-\phi^d_q)} 
\left[ (\Vt_0)^*_{tb} (\Vt_0)_{tq} \right]^2\,,\nonumber\\
\mathcal{C}_1^{LR}(\mu_R,B_q)&=& - c_{LR} y_t^4  e^{i(\phi^d_3-\phi^d_q)} V^*_{tb} V_{tq}(\Vt_0)^*_{tb} (\Vt_0)_{tq}~.
\end{eqnarray}
Working in the limit $\tc_{12} \ll 1$ (hence $\ts_{12} \approx 1$) 
we get
\begin{eqnarray}
&& \mathcal{C}_1^{VRR}(\mu_R,B_d) \approx -\frac{c_{RR}}{4} y_t^4 
e^{2i(\phi^d_3-\phi^d_1)} \tc_{12}^2~, \qquad 
\mathcal{C}_1^{VRR}(\mu_R,B_s) \approx - \frac{c_{RR}}{4} y_t^4 
e^{2i(\phi^d_3-\phi^d_2)}  \,,\nonumber\\
&& \mathcal{C}_1^{LR}(\mu_R,B_d) \approx \mp \frac{c_{LR}}{2} y_t^4  
e^{i(\phi^d_3-\phi^d_1)} V^*_{tb} V_{td}  \tc_{12}~, \qquad
\mathcal{C}_1^{LR}(\mu_R,B_s) \approx \mp  \frac{c_{LR}}{2}  y_t^4
e^{i(\phi^d_3-\phi^d_2)} V^*_{tb} V_{ts}~, \nonumber \\
&& \mathcal{C}_1^{VLL}(\mu_R,B_{s,d})=0\,, \label{eq:93ab}
\end{eqnarray}
where the $\mp$ sign reflects the  $\pm$ in (\ref{eq:Vtansatz2}).
Using these expressions in the formulae for $M^q_{12}$ we obtain
\bea
&& (M^d_{12})_{\rm SM+RH} =  (M^d_{12})_{\rm SM} \times\Big[ 1 + 
\no \\ && \qquad
\left( -6.1\times10^{3} \times  c_{RR} \tc_{12}^2 e^{-2i(\phi^d_{31}+\beta)}
\pm 4.7 \times 10^{2} \times c_{LR} \tc_{12} e^{-i(\phi^d_{31}+\beta)} 
\right) \frac{(3~{\rm TeV})^2}{\Lambda^2}~\Big]~,\qquad  \\
&& (M^s_{12})_{\rm SM+RH} =  (M^s_{12})_{\rm SM} \times\Big[ 1 + 
\no \\ && \qquad
\left(-2.5\times10^{2} \times  c_{RR}  e^{-2i(\phi^d_{32}+\beta_s)} 
\mp 0.9 \times10^{2} \times c_{LR} e^{-i(\phi^d_{32}+\beta_s)} \right) 
\frac{(3~{\rm TeV})^2}{\Lambda^2}~\Big]~,
\eea
where, similarly to $\phi^d_{21}$,  we have defined $\phi^d_{3i}=(\phi^d_3-\phi^d_i)$.
In obtaining the numerical values we have evaluated  $(M^q_{12})_{\rm SM}$
using the inputs in Table~\ref{tab:num}.

As anticipated, given the bounds on $c_{RR} \tc_{12}^2$ and $c_{LR}\tc_{12}$
in (\ref{eq:epsbound1})--(\ref{eq:epsbound2})
following from the neutral kaon system, in this framework non-standard 
contributions to both modulo and phase of $B_d$ mixing are safely negligible.
On the other hand, sizable contributions to the $B_s$ system are possible 
if $c_{RR,LR}$ are in the $10^{-3}$--$10^{-2}$ range and $\tc_{12}$ is small 
enough to satisfy the kaon bounds.

If we do not assume $\tc_{12} \ll 1$ and, more generally, 
go beyond the ansatz (\ref{eq:Vtansatz2}) for the RH mixing 
matrix, sizable contributions to $B_d$ mixing are possible  
assuming $\ts_{12} \ll 1$. However, as already stated, 
this precludes the possibility of large effects in the $B_s$ system.

\subsection{\boldmath Combined fit of $\varepsilon_K$ and $B_s$ mixing}
\label{section:DFglobal}
Here we discuss in more detail the interesting scenario with $c_{RR,LR}$
in the $10^{-3}$--$10^{-2}$ range
and $c_{12} \ll 1$, where we can accommodate a large CP-violating phase 
in $B_s$ mixing, as hinted by CDF~\cite{Aaltonen:2007he} 
and D0~\cite{Abazov:2008fj,Abazov:2010hv} and, at the same time, 
satisfy the bounds from $\varepsilon_K$.
Values of $c_{RR,LR}$ in the $10^{-3}$--$10^{-2}$ range
are substantially lower than 
the $\cO(1)$ Wilson coefficients determined from charged-current 
interactions (assuming $\Lambda=3$~TeV as reference scale).
However, it is perfectly conceivable that the $\Delta F=2$ operators 
are loop-suppressed with respect to the charged-current ones, such that 
a $10^{-3}$--$10^{-2}$ relative suppression of the 
corresponding Wilson coefficients can naturally be accommodated. 

We stress that a large CP-violating phase in $B_s$ mixing
is not a clear prediction of the model we are considering. 
As pointed out in the previous section,
the only clear prediction is the absence of 
significant contributions to $B_d$ mixing,
 implied by the 
bounds from the kaon system, 
if we require a  
large CP-violating phase in $B_s$ mixing.
Still, it is interesting to check if 
this experimental ``anomaly'' can be solved with
reasonable values of the free parameters of the effective theory
we are considering. To better investigate this point, we analyse 
separately the cases where the leading correction to the SM in $B_s$ mixing 
is induced by $c_{RR}$ and $c_{LR}$, respectively.

Assuming $c_{LR} \ll c_{RR}$ and neglecting the tiny contribution of $\beta_s$,
the constraints of the $B_s$ system imply
\bea
&& \frac{ (\Delta M_s)_{\rm SM+RH} }{ (\Delta M_s)_{\rm SM} }  =  
\left| 1 -2.6\times10^{2} \times  c_{RR}  e^{-2i \phi^d_{32} } \right| 
= \frac{ (\Delta M_s)_{\rm exp} }{ (\Delta M_s)_{\rm SM} } 
~\approx~ 0.96 \pm 0.15~,\qquad   \\
&& S_{\psi \phi } = - \frac{ 2.6\times10^{2} \times  c_{RR}  \sin( 2 \phi^d_{32}) }{
\left| 1 -2.6\times10^{2} \times  c_{RR}  e^{-2i \phi^d_{32} } \right| }
~\approx~ 0.6 \pm 0.3~,
\eea 
where the latter numerical entry is only a rough indicative value 
for the CP asymmetry favored by the Tevatron experiments
(for more details see~\cite{Aaltonen:2007he,Abazov:2008fj,Abazov:2010hv}
and the model-independent analysis in \cite{Ligeti:2010ia}). 
The central values of these equations are full-filled with 
the following four-fold solution:
\bea
&& c_{RR} \approx \pm 7.3 \times 10^{-3} \quad {\rm and} \quad \sin(2\phi^d_{32}) \approx \mp 0.30~, \no \\
&& c_{RR} \approx \pm 2.3 \times 10^{-3} \quad {\rm and} \quad \sin(2\phi^d_{32}) \approx \mp 0.95~.
\label{eq:sinphi32_a}
\eea
As anticipated, these values are in good agreement with the na\"ive expectation 
of a $1/(16\pi^2) \approx 6 \times 10^{-3}$ suppression
between this $\Delta F=2$ operator (presumably generated at the loop level)
relative to those contributing to right-handed charged currents 
(presumably generated at the tree level).

Having fixed $c_{RR}$, we can ask which are the values of the mixing angles
necessary to satisfy the bounds from the kaon system. 
To this end, it is first important to note that, due to the higher value 
of $|V_{ub}|$ determined from charged-current processes with 
the inclusion of RH currents, the prediction of $\varepsilon_K$ in this
framework {\em without} extra contributions is in excellent agreement with 
data (contrary to what happens in the SM~\cite{Lunghi:2008aa,Buras:2008nn}). 
Indeed the value of $\sin(2\beta)$ determined by tree-level
observables only, namely $|V_{ub}|$ and $\gamma$ (following the 
analysis in~\cite{Bona:2007vi}) is
\be
\sin(2\beta)^{\rm RH}_{\rm tree} = 0.77 \pm 0.05~.
\ee
This value is substantially higher than the corresponding result obtained 
in the SM, $\sin(2\beta)^{\rm SM}_{\rm tree} = 0.734 \pm 0.034$~\cite{Bona:2007vi},
where the inclusive and exclusive determinations of $|V_{ub}|$ 
are averaged.  As a result of this higher value of $\sin(2\beta)$,
the tension between the experimental value of $\varepsilon_K$ 
and its prediction within the SM (see e.g.~\cite{Brod:2010mj,Buras:2010mh} for a recent
analysis) is automatically solved. 

Despite there is no need for non-standard contributions to  $\varepsilon_K$,
the theoretical errors on this observable allow for extra contributions 
within $\approx \pm 10\%$ of the SM amplitude. 
This condition is obtained for 
\bea 
&& |\tc_{12}| |\sin(2\phi^d_{21}) |^{1/2} < 1.9 \times 10^{-3}~,
\qquad {\rm for}~|c_{RR}| \approx 7.3 \times 10^{-3}~, \no \\ 
&& |\tc_{12}| |\sin(2\phi^d_{21}) |^{1/2} < 3.4 \times 10^{-3}~, 
\qquad {\rm for}~|c_{RR}| \approx 2.3 \times 10^{-3} ~.
\label{eq:c21fromRR}
\eea
These values are small but not highly fine-tuned: for CP-violating phases of $\cO(0.1)$,
the mixing angle $\tc_{12}$ can reach values of $\cO(10^{-2}$), which are 
larger than the CKM element $|V_{ub}|$.

Due to the large chiral enhancement of the contribution of $\mathcal{Q}_1^{LR}$
to $\varepsilon_K$, more fine-tuning is required if $c_{LR}$ provides the dominant
contribution to $B_s$ mixing. Indeed repeating the above argument for $c_{RR} \ll c_{LR}$ 
leads to the following four-fold solution from $B_s$ mixing,
\bea
&& c_{LR} \approx \pm 2.0 \times 10^{-2} \quad {\rm and} \quad \sin(2\phi^d_{32}) \approx \mp 0.30~, \no \\
&& c_{LR} \approx \pm 0.6 \times 10^{-2} \quad {\rm and} \quad \sin(2\phi^d_{32}) \approx \mp 0.95~,
\label{eq:sinphi32_b}
\eea
and the following conditions from $\varepsilon_K$:
\bea
&& |\tc_{12} \sin(\phi^d_{21}-\beta+\beta_s) |  < 0.2 \times 10^{-4}~,
\qquad {\rm for}~|c_{LR}| \approx 2.0 \times 10^{-2}~, \no \\ 
&& |\tc_{12} \sin(\phi^d_{21}-\beta+\beta_s) |  < 0.6 \times 10^{-4}~, 
\qquad {\rm for}~|c_{LR}| \approx 0.6 \times 10^{-2} ~.
\label{eq:c21fromLR}
\eea
It is clear that in this case the condition on the 1-2 mixing 
is more stringent than in (\ref{eq:c21fromRR}).

\subsection{Summary}
The main results of the $\Delta F=2$ analysis can be summarized as follows:
\begin{itemize}
\item{}
A large CP-violating phase in $B_s$ mixing, as hinted by the 
Tevatron experiments, 
can be accommodated for natural values of the free parameters. 
The two necessary ingredients for this mechanism to work are: 
1) Wilson coefficients of $\cO(1/(16\pi^2))$ for the $\Delta F=2$ operators 
in (\ref{eq:DF2effRH}), assuming as reference scale $\Lambda =3$~TeV; 
2) a RH mixing matrix with the following from 
\be
\left| \Vtr^{(B_s~{\rm mixing})} \right|  \sim
\left(\begin{array}{ccc}
   0  & \frac{\sqrt{2}}{2} &  \frac{\sqrt{2}}{2} \\ 
   1  &  0  &  0 \\ 
   0  & \frac{\sqrt{2}}{2} &  \frac{\sqrt{2}}{2}
\end{array}\right)~,
\label{eq:VtansatzBs}
\ee
where the null entries should not be taken as exact zeros, 
but rather as very small entries. 
\item{}
According to the RH mixing structure in (\ref{eq:VtansatzBs}),
with improved experimental precision it should be possible 
to resolve the presence of RH currents in $s\to u$ 
charged-current transitions, or a $\cO(10^{-3})$ 
deviation in the determination of $|V_{us}|$ from $K\to \pi\ell\nu$
and $K\to \ell\nu$ decays.
\item{}
Thanks to the large value of $\sin(2\beta)$, 
following from the inclusion of RH currents in the determination 
on $|V_{ub}|$, the prediction of $\varepsilon_K$ in this
framework is in excellent agreement with data
without extra contributions.
\item{} 
The combination of a large CP-violating phase in $B_s$ mixing
and only small NP effects allowed by $\varepsilon_K$ implies 
negligible effects in $B_d$ mixing. This, in turn, implies a tension 
between the measured valued of $S_{\psi K}$ and the predicted 
value of $\sin(2\beta)$ in this framework. The only possibility to 
solve this problem is to assume $\ts_{12} \ll 1$, giving up the 
possibility of NP effects in  $B_s$ mixing.
\end{itemize}

\section{\boldmath $Z$-mediated FCNCs and $Z\to b\bar b$}

\subsection{Modification of the  RH couplings of the $Z$ boson}

As discussed in Section~\ref{sect:d6fcnc}, in the $\Delta F=1$ sector 
we focus our attention on the effective Lagrangian (\ref{eq:LZgen}).
The effective operators appearing in this Lagrangian are equivalent 
to $O^{(6)}_{R_{h1}}$ and $O^{(6)}_{R_{h2}}$ analysed in Section~\ref{sect:cc1}, 
but for the additional insertion of the combination 
of Yukawa matrices $(\YU^\dagger \YU)_{ij}$.
After the breaking of the electroweak symmetry, 
they lead to the following effective
right-handed couplings of the $Z$ boson to down-type quarks:
\be
\cL_{\rm eff}^{(Z_R)} =  -  \frac{g}{c_W}  
 \frac{v^2(c_{R_{Z1}} + 2 c_{R_{Z2}})  }{2 \Lambda^2}  
y_t^2 (\Vt_{ti}^* \Vt_{tj}) \bar d^i_R \gamma^\mu d^j_R  Z_\mu~,
\label{eq:ZR}
\ee
where $c_W=\cos\Theta_W$ (similarly, in the following 
we use $s_W=\sin\Theta_W$).

Denoting the effective couplings of the $Z$ 
to down-type quarks as follows
\be
\cL_{\rm eff}^{Z} =   \frac{g}{ c_W} \left( ~g_L^{ij}~\bar d^i_L \gamma^\mu d^j_L +
 g_R^{ij}~\bar d^i_R \gamma^\mu d^j_R \right) Z_\mu~,
\ee
the SM contribution, evaluated at the one-loop level in the 't Hooft-Feynman gauge
in the large top-mass limit, is 
\bea
(g_L^{ij})_{\rm SM} &=& \left(-\frac{1}{2}+\frac{1}{3} s_W^2\right)\delta_{ij} +
\frac{g^2}{8\pi^2} V_{ti}^* V_{tj} C_0(x_t)~, \qquad x_t=\frac{m_t^2}{m_W^2}~,
 \\
(g_R^{ij})_{\rm SM} &=& \frac{1}{3} s_W^2 \delta_{ij}~.
\eea
The loop function $C_0(x_t)$, that  in the large $x_t$ limit
is gauge independent ($g^2C_0(x_t)\to g^2x_t/8 = y^2_t/4$ for $m^2_t \gg m_W^2$),
can be found in \cite{Buchalla:1995vs}. 
Using these notations, the effect of the RH operators  $O^{(6)}_{R_{h1}}$ and $O^{(6)}_{R_{h2}}$ 
can be included as a modification of the RH effective coupling:
\be
(g^{ij}_R)_{\rm tot} = (g^{ij}_R)_{\rm SM} +(\Delta g_R^{ij})_{RH}~, 
\qquad
(\Delta g_R^{ij})_{RH} =  -   
 \frac{v^2(c_{R_{Z1}} + 2 c_{R_{Z2}})  }{2 \Lambda^2}  
y_t^2 \Vt_{ti}^*\Vt_{tj}.
\label{eq:gRnew}
\ee

\subsection{\boldmath $Z\to b\bar b$}
\label{sect:Zbb}
The experimental determination of the effective couplings of the $Z$
bosons to $b$ quarks resulting from the global fit of 
electroweak data collected by the LEP and the SLD experiments
is~\cite{:2005ema}
\bea
(g_L^{bb})_{\rm exp} &=& -0.4182 \pm 0.0015~,  \\
(g_R^{bb})_{\rm exp} &=& +0.0962 \pm 0.0063~.  
\eea
While the result for the LH coupling is consistent with the 
SM prediction, there is a large disagreement between data and 
SM expectation in the RH sector:
\be
(\Delta g_R^{bb})_{\rm exp}  
 =  (g_R^{bb})_{\rm exp} -  (g_R^{bb})_{\rm SM} = (1.9 \pm 0.6) \times 10^{-2}~.
\label{eq:Zbb_exp}
\ee 
This deviation could in principle be solved by choosing appropriate couplings 
for the RH operators in (\ref{eq:Zops}). From the modified RH 
coupling in (\ref{eq:gRnew}) we get
\be
(\Delta g_R^{bb})_{RH}  \approx  -  0.15 \times 10^{-2} \times 
c_{Z_R}^{\rm eff}
\label{eq:gbbRH}
\ee
where 
\be
c_{Z_R}^{\rm eff}
 = (c_{R_{Z1}} + 2 c_{R_{Z2}})  \frac{(3~{\rm TeV})^2}{\Lambda^2}~,
\label{eq:CZeff}
\ee
and the numerical value has been obtained assuming $|\Vt_{tb}|^2\approx 1/2$
(see Sect.~\ref{sect:globalfit}). As can be seen, for $\Lambda =3$~TeV  
and $c_{R_{Zi}}=\cO(1)$, the correction is too small 
to contribute significantly to the experimental discrepancy in 
(\ref{eq:Zbb_exp}). In principle, the effect could be explained assuming 
 $\Lambda =1$~TeV and $c_{R_{Z1}},c_{R_{Z2}}=\cO(1)$. However,
as we will show in the following, this possibility is ruled out 
after taking into account the 
phenomenological bounds from rare $B$ decays.

\subsection{\boldmath Rare $B$ and $K$ decays: preliminaries}
General phenomenological analyses about the role of $Z$-mediated 
RH currents in $B$ and $K$ decays with 
a lepton pair in the final state can be found 
in~\cite{Buchalla:2000sk,Altmannshofer:2009ma}.
While most of the results obtained in these two papers
can be applied also to the present study, here we 
go one step forward having determined a series of 
constraints on the flavour structure of the 
RH mixing matrix from other processes. We begin this section by
introducing the relevant effective Hamiltonians.

For $B_{s,d}\to \mu^+\mu^-$ channels we generalize the SM effective Hamiltonian to 
\bea
&& \Heff =
-\frac{4 G_F}{\sqrt{2}} \frac{\alpha}{2\pi s^2_W} V_{tb}^\ast 
V_{tq} \times \left[
Y_{LL} (\bar b_L \gamma^\mu  s_L)(\bar \mu_L \gamma_\mu \mu_L)
\right. \no  \\  
&& \quad 
\left.
 + Y_{LR} (\bar b_L \gamma^\mu  s_L)(\bar \mu_R \gamma_\mu \mu_R)
 +Y_{RL} (\bar b_R \gamma^\mu s_R)(\bar \mu_L \gamma_\mu \mu_L)
 +Y_{RR} (\bar b_R \gamma^\mu s_R)(\bar \mu_R \gamma_\mu \mu_R) \right]\,, \qquad\quad
\label{eq:heffBmumu}
\eea
where $q=d,s$. The overall factor allows for an easy comparison 
with the SM, where ${Y_{LL}-Y_{LR}}=Y_0(x_t)$ and $Y_{RR}=Y_{RL}=0$.
The contributions to the $Y$ functions with RH 
quark currents obtained 
by means of the effective Lagrangian (\ref{eq:ZR}) are
\bea
Y_{RL}-Y_{RR} =  - T~\frac{ \Vt_{tb}^* \Vt_{tq} }{ V_{tb}^* V_{tq} }~, \qquad
Y_{RL}+Y_{RR} =  -(1-4 s_W^2)~T~\frac{ \Vt_{tb}^* \Vt_{tq} }{ V_{tb}^* V_{tq} }~,
\label{eq:YRRZ} 
\eea
where we have defined 
\be
T = (c_{R_{Z1}} + 2 c_{R_{Z2}})
\frac{4\pi^2 v^2 y_t^2}{g^2 \Lambda^2} = 
0.55 \times \left(\frac{m_t(m_t)}{163.5~{\rm GeV}}\right)^2  c_{Z_R}^{\rm eff}~.
\ee
Note that $Y_{LL}$ and $Y_{LR}$ are not affected 
by the effective Lagrangian (\ref{eq:ZR}).

A simple generalization of the SM effective Hamiltonians 
can be implemented also to describe $K \to \pi \nu\bar \nu$ 
and  $B \to \{ X_s,K, K^*\} \nu\bar \nu$ 
decays, with the further simplification that we can neglect 
operators with $\nu_R$ fields, that we assumed to be heavy. 
In the $K \to \pi \nu\bar \nu$ case we generalize 
the short-distance effective Hamiltonian to 
\be
\Heff =
\frac{4 G_F}{\sqrt{2}} \frac{\alpha}{2\pi s^2_W} V_{ts}^\ast 
V_{td} \times \left[ X_{LL}(K) (\bar s_L \gamma^\mu  d_L) 
 +X_{RL}(K) (\bar s_R \gamma^\mu d_R)\right] \times (\bar \nu_L \gamma_\nu \nu_L)\,,
\label{eq:heffKnn}
\ee
where the leading SM top-quark contribution yields $X_{RL}=0$
and $X_{LL} \equiv X_{\rm SM} =1.464 \pm 0.041$~\cite{Buras:2006gb}
(for simplicity we omit the sub-leading charm contribution
that will be included in the phenomenological analysis).
With these notations the non-standard contribution  
obtained by means of the effective Lagrangian (\ref{eq:ZR}) is
\be
X_{RL}(K)  =  - T~\frac{ \Vt_{ts}^* \Vt_{td} }{ V_{ts}^* V_{td} }~.
\label{eq:XRL}
\ee
In the $B \to  \{ X_s, K, K^*\} \nu\bar \nu$ case the 
general effective Hamiltonian is 
\be
\Heff =
\frac{4 G_F}{\sqrt{2}} \frac{\alpha}{2\pi s^2_W} V_{tb}^\ast 
V_{ts} \times \left[ X_{LL}(B_s) (\bar b_L \gamma^\mu  s_L) 
 +X_{RL}(B_s) (\bar b_R \gamma^\mu s_R)\right] \times (\bar \nu_L \gamma_\nu \nu_L)\,,
\label{eq:heffBXsnn}
\ee
with 
\be
X_{RL}(B_s)  =  - T~\frac{ \Vt_{tb}^* \Vt_{ts} }{ V_{tb}^* V_{ts} }~.
\label{eq:XRLBs}
\ee

\subsection{\boldmath $B_{s,d} \to \mu^+\mu^-$}

When evaluating the amplitude for  $B_s\to \mu^+\mu^-$  
by means of (\ref{eq:heffBmumu}) the following simplifications occur
\be
\langle 0| \bar b \gamma_\mu P_{R,L} s | B^0\rangle =
 \pm \frac{1}{2} \langle 0| \bar b \gamma_\mu \gamma_5 s | B^0\rangle~,
\qquad 
\langle \bar\mu \mu| \bar \mu \gamma_\mu P_{R,L} \mu |  0  \rangle =
\pm \frac{1}{2} \langle \bar\mu \mu| \bar \mu \gamma_\mu \gamma_5 \mu | 0 \rangle\,.
\ee
The resulting branching ratio
is then obtained from the known SM expression (see e.g.~\cite{Buras:2003td})
by making the following replacement 
\begin{equation}
Y_0(x_t)\to Y_{LL}+Y_{RR}-Y_{RL}-Y_{LR} \equiv Y_{tot}
\end{equation}
so that 
\begin{equation}
\cB(B_s\to \ell^+\ell^-) = \tau(B_s)\frac{G^2_{\rm F}}{\pi}
\left(\frac{\alpha}{4\pi s^2_W}\right)^2 F^2_{B_s}m^2_l m_{B_s}
\sqrt{1-4\frac{m^2_l}{m^2_{B_s}}} |V^\ast_{tb}V_{ts}|^2 |Y_{tot}|^2\,.
\end{equation}
The expression for $\cB(B_d\to \ell^+\ell^-)$ is obtained by replacing $s$ by $d$. 

Taking into account that $\Vt_{tb}^* \Vt_{td} \approx \pm \tc_{12} e^{i\phi_{31}^d}/2$  
and  $\Vt_{tb}^* \Vt_{ts} \approx \pm \ts_{12}  e^{i\phi_{32}^d}/2$ 
(see Sect.~\ref{sect:globalfit} and Sect.~\ref{sect:globalfit}), 
and using (\ref{eq:YRRZ}), we finally obtain the following expressions 
for the two branching ratios normalized to the SM:
\bea
&&\!\! \cB(B_s\to \ell^+\ell^-) = \cB(B_s\to \ell^+\ell^-)_{\rm SM} 
\left| 1 \mp  7.8 \times \ts_{12}  e^{i\phi_{32}^d}~c_{Z_R}^{\rm eff}  \right|^2, \no \\
&&\!\! \cB(B_d\to \ell^+\ell^-) = \cB(B_d\to \ell^+\ell^-)_{\rm SM} 
\left| 1 \pm  37 \times  \tc_{12}  e^{i\phi_{31}^d}~c_{Z_R}^{\rm eff} \right|^2~. \qquad\ 
\label{eq:BRllRH}
\eea
The muon channels are those where the experimental searches are closer to 
the SM predictions. The numerical values of the latter, obtained using 
the relation of ${\cB}(B_q\to\mu^+\mu^-)$ to $\Delta M_q$ pointed out 
in~\cite{Buras:2003td}, are
\be\label{TH}
\cB(B_s\to \mu^+\mu^-)= (3.2\pm0.2)\times 10^{-9}~, \qquad
\cB(B_d\to \mu^+\mu^-)= (1.0\pm0.1)\times 10^{-10}~.
\ee
These figures should be compared with the $95\%$ C.L. upper limits from 
CDF  \cite{Aaltonen:2007kv} and D0 \cite{Abazov:2007iy} (in parentheses)
\be\label{CDFD0}
\cB(B_s\to \mu^+\mu^-)\le 3.3~(5.3)\times 10^{-8}, \qquad
\cB(B_d\to \mu^+\mu^-)\le 1 \times 10^{-8}.
\ee
Using the results in (\ref{eq:BRllRH}) these limits imply 
\be
 \left| \ts_{12} c_{Z_R}^{\rm eff} \right| < 0.54~, \qquad 
 \left| \tc_{12} c_{Z_R}^{\rm eff} \right| < 0.30~,
\ee
where the bounds have been derived taking into 
account the interference with the SM (and choosing the maximal
interference effect). These two limits can be combined to derive  
the following bound 
\be
 \left|  c_{Z_R}^{\rm eff} \right| < 0.62~,
\label{eq:boundBll}
\ee
which holds independently of any assumption about the value of $\tc_{12}$.
Using this bound in (\ref{eq:gbbRH}) we get
\be
\left|(\Delta g_R^{bb})_{RH} \right| < 1\times 10^{-3}~,
\ee 
which, by construction, does not rely on any assumption about the 
value of $\tc_{12}$.
It is then clear that, despite the presence of a non-standard 
coupling of the $Z$ boson to RH fermions, within our effective theory 
the constraints from  $\cB(B_{d,s}\to \ell^+\ell^-)$ prevent 
a solution to the $Z \to b \bar b$ anomaly in (\ref{eq:Zbb_exp}).

The bound (\ref{eq:boundBll}) has been derived from the experimental 
bounds on both $\cB(B_{s}\to \mu^+\mu^-)$ and  $\cB(B_{d}\to \mu^+\mu^-)$ 
in order to show in a simple manner that there is no room for a 
sizable contribution to $Z \to b \bar b$ in our framework, 
independently of the value of  $\tc_{12}$.  
As far as the maximal enhancement of $\cB(B_{s}\to \mu^+\mu^-)$ is concerned, 
a more stringent bound can be derived taking into account the 
constraints on the effective $Z^\mu \bar b_R \gamma^\mu s_R$ 
coupling following 
from $\cB(B_{s,d}\to X_s \ell^+\ell^-)$~\cite{Buchalla:2000sk,Altmannshofer:2009ma}. 
In particular, from the bound reported in \cite{Altmannshofer:2009ma}
one finds
\be
| T | 
\times \left|  \frac{ \Vt_{tb}^* \Vt_{ts} }{ V_{tb}^* V_{ts} } \right| < 1.07~,
\ee
at the $90\%$ C.L.~level, or
\be
 \left| \ts_{12}  c_{Z_R}^{\rm eff} \right| < 0.15~.
\label{eq:Zbs_lastb}
\ee
Using this result in (\ref{eq:BRllRH}), the maximal 
enhancement in $\cB(B_{s}\to \mu^+\mu^-)$ over its 
SM expectation does not exceed a factor of 5.
This should be contrasted to other NP frameworks, 
in particular to models with non-standard scalar FCNCs,
where the present experimental upper bound on 
 $\cB(B_{s}\to \mu^+\mu^-)$ could easily be saturated. 

In case of an $\cO(1)$ deviation from the SM in  $\cB(B_{s}\to \mu^+\mu^-)$,
a clear prediction of our framework, following from the analysis of 
$\Delta F=2$ processes, is the absence of visible deviations
from the SM in $\cB(B_{d}\to \mu^+\mu^-)$. Indeed, as we have seen 
in Section~\ref{section:DFglobal}, 
the configuration of the RH matrix that could allow to 
explain a large $S_{\psi\phi}$ asymmetry requires  $\ts_{12} \approx 1$ and 
$\tc_{12} < 10^{-2}$. When combined with the bound in 
(\ref{eq:Zbs_lastb}) this condition implies negligible non-standard 
effects in  $\cB(B_{d}\to \mu^+\mu^-)$.

\subsection{\boldmath $B \to \{X_s,K, K^*\} \nu\bar \nu$}

Following the analysis of Ref.~\cite{Altmannshofer:2009ma}, 
the branching ratios of the $B \to \{X_s,K, K^*\}\nu\bar \nu$  
modes in the presence of RH currents can be written as follows
\bea
\cB(B\to K \nu \bar \nu) &=& 
\cB(B\to K \nu \bar \nu)_{\rm SM} \times\left[1 -2\eta \right] \epsilon^2~, \label{eq:BKnn}\\
\cB(B\to K^* \nu \bar \nu) &=& 
\cB(B\to K^* \nu \bar \nu)_{\rm SM}\times\left[1 +1.31\eta \right] \epsilon^2~, \\
\cB(B\to X_s \nu \bar \nu) &=& 
\cB(B\to X_s \nu \bar \nu)_{\rm SM} \times\left[1 + 0.09\eta \right] \epsilon^2~,\label{eq:Xsnn}
\eea
where we have introduced the variables 
\be
\epsilon^2 = \frac{ |X_{\rm LL}|^2 + |X_{\rm RL}|^2 }{ 
|X_{\rm LL}^{\rm SM}|^2 }~,  \qquad
\eta = \frac{ - {\rm Re} \left( X_{\rm LL}^* X_{\rm RL}\right) }{ |X_{\rm LL}|^2 + |X_{\rm RL}|^2 }~, 
\ee
in terms of the Wilson coefficient of the effective 
Hamiltonian (\ref{eq:heffBXsnn}).\footnote{
The expressions in Eqs.~(\ref{eq:BKnn})--(\ref{eq:Xsnn}), 
as well as the SM figures in (\ref{eq:BKnnSM}), refer only to the short-distance contributions 
to these decays. The latter are obtained from the corresponding total rates 
subtracting the reducible long-distance effects pointed out in~\cite{Kamenik:2009kc}.}
To simplify the notations, here and in the following 
we omit to specify the meson system in the $X_{LL,LR}$ functions. 
The updated predictions for the SM branching  ratios 
are~\cite{Bartsch:2009qp,Kamenik:2009kc,Altmannshofer:2009ma}
\bea
\cB(B\to K \nu \bar \nu)_{\rm SM}   &=& (3.64 \pm 0.47)\times 10^{-6}~, \no \\
\cB(B\to K^* \nu \bar \nu)_{\rm SM} &=& (7.2 \pm 1.1)\times 10^{-6}~, \no \\
\cB(B\to X_s \nu \bar \nu)_{\rm SM} &=& (2.7 \pm 0.2)\times 10^{-5}~, 
\label{eq:BKnnSM}
\eea
to be compared with the experimental bounds~\cite{Barate:2000rc,:2007zk,:2008fr}
\bea
\cB(B\to K \nu \bar \nu)   &<&  1.4 \times 10^{-5}~, \no \\
\cB(B\to K^* \nu \bar \nu) &<&  8.0 \times 10^{-5}~, \no \\
\cB(B\to X_s \nu \bar \nu)  &<&  6.4 \times 10^{-4}~.
\label{eq:BKnn_exp}
\eea

%

The expressions in Eqs.~(\ref{eq:BKnn})--(\ref{eq:Xsnn}) 
are valid for wide class of NP model: all models
giving rise to the effective Hamiltonian (\ref{eq:heffBXsnn}). 
In our specific framework NP effects are encoded 
only in the RH sector and the $X_{RL}$ function is given in~(\ref{eq:XRLBs}).
The variables $\epsilon$ and $\eta$ then assume the following form:
\bea
\epsilon^2 &=&  1 +  \frac{T^2}{ X^2_0(x_t) } 
\left| \frac{ \Vt_{tb}^* \Vt_{ts} }{ V_{tb}^* V_{ts} }\right|^2  
\approx 1 + 22.1 \times |\ts_{12} c_{Z_R}^{\rm eff}|^2 ~,  \\
\eta &=& \frac{T}{\epsilon^2 X_0(x_t)} {\rm Re} \left( \frac{ \Vt_{tb}^* \Vt_{ts} }{ V_{tb}^* V_{ts} }
 \right) 
\approx  \mp \frac{ 4.7 \times \ts_{12} \cos(\phi_{32}^d)  c_{Z_R}^{\rm eff} }{ 1 + 22.1 \times 
|\ts_{12} c_{Z_R}^{\rm eff}|^2}~.
\eea

Taking into account the bound on $\ts_{12} c_Z^{\rm eff}$ in Eq.~(\ref{eq:Zbs_lastb}), we find 
that the predictions for the exclusive branching ratios can be 
enhanced by more than a factor of two over the corresponding SM estimates.
On the contrary, the enhancement in the inclusive mode does not exceed $50\%$. 
Most important, a clear prediction of RH currents is the anti-correlation of the two 
exclusive modes: if  $\cB(B\to K \nu \bar \nu)$ is enhanced then 
$\cB(B\to K^* \nu \bar \nu)$ is suppressed, and viceversa.\footnote{In principle, 
similar correlations could also be established in the 
$B\to K \ell^+\ell^-$ and $B\to K^* \ell^+ \ell^-$ channels.
However, in this case the pattern is less clean due to the presence of 
other effective operators.
Moreover, the simplifying assumption of considering only 
the effective Lagrangian (\ref{eq:LZgen})
as representative of the dominant RH effects 
is not necessarily a good approximation for these channels.
A detailed analysis of $B\to K \ell^+\ell^-$ and $B\to K^* \ell^+ \ell^-$ 
in our effective theory goes beyond the purpose of the present paper
and we refer to the general model-independent analysis 
in Ref.~\cite{Altmannshofer:2008dz}.}  
Not surprisingly, the pattern of these three modes is 
very similar to what observed in the three modes 
relevant for the determination of $|V_{ub}|$ in Section~\ref{sect:ccbounds}.
However, in the rare modes the deviations from the SM can in principle 
be larger than in the charged-current decays. 

\begin{figure}[t]
\begin{center}
\includegraphics[width=.6\textwidth]{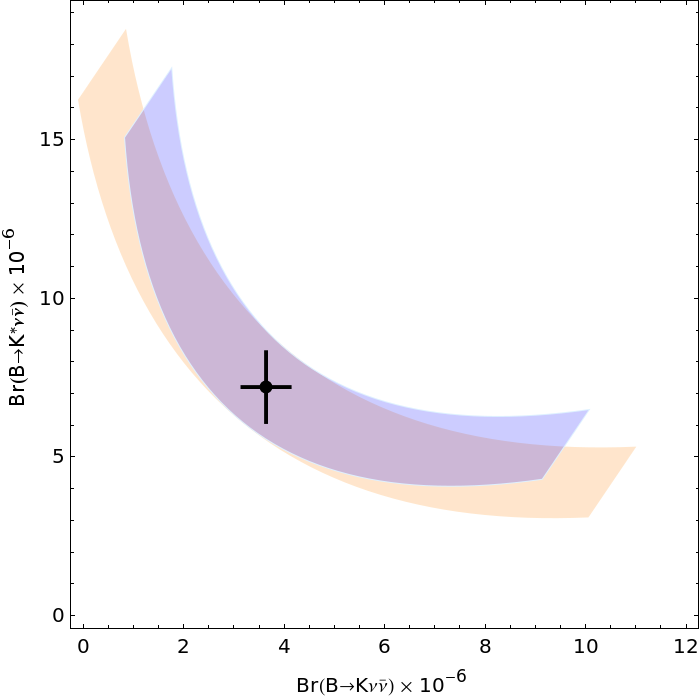}
\end{center}
\caption{\label{fig:BKnn} 
Correlation between
$\cB(B\to K \nu \bar \nu)$ and 
$\cB(B\to K^* \nu \bar \nu)$ in our effective theory.
The two bands correspond to the two values of $|\sin(2\phi^d_{32})|$ 
in Eqs.~(\ref{eq:sinphi32_a}), with the uncertainty given by the
errors on the SM predictions:
blue (dark gray) band for $|\sin(2\phi^d_{32})|=0.95$, 
orange (light gray) band for $|\sin(2\phi^d_{32})|=0.30$. 
The black point denotes
the SM values with the corresponding error bars. 
}
\end{figure}

The correlations among the three $B \to \{X_s,K, K^*\} \nu\bar \nu$ modes 
are in principle affected by the uncertainty on the CP-violating 
phase $\phi_{32}^d$. However, the same phase enters 
in $S_{\psi\phi}$. If we require a large $S_{\psi\phi}$ (as hinted by CDF and D0), 
this uncertainty is strongly reduced, as shown in  Figure~\ref{fig:BKnn}.
In this plot we show the expectations of the two exclusive branching ratios 
for the preferred values of the CP-violating phase $\phi_{32}^d$ 
as determined from $S_{\psi\phi}$ in Section~\ref{section:DFglobal}.
Two points should be noted: 1) only the modulo of $\sin(2\phi^d_{32})$
enters in the branching ratios of the rare modes 
(via their $\cos(\phi^d_{32})$ dependence), as a result, 
there are only two independent choices corresponding to all 
the solutions in in Eqs.~(\ref{eq:sinphi32_a}) 
and Eqs.~(\ref{eq:sinphi32_b}); 2) these two choices give rise to 
predictions for the $B \to \{K, K^*\} \nu\bar \nu$ branching 
ratios which are almost indistinguishable. As shown in 
Figure~\ref{fig:BKnn}, the correlation pattern 
is very clean and, if observed, would provide a clear confirmation 
of this framework.

\subsection{\boldmath $K \to \pi \nu\bar \nu$}

The SM branching ratios for the two most interesting 
$K \to \pi \nu\bar \nu$ modes can be written 
as~\cite{Buras:2005gr,Isidori:2005xm,Mescia:2007kn,Brod:2008ss}
\begin{gather} \label{eq:BRSMKp}   
  \cB (K^+\to \pi^+ \nu\bar\nu) = \kappa_+ \left [ \left ( \frac{{\rm Im} X_{\rm eff} }{\lambda^5}
  \right )^2 + \left ( \frac{{\rm Re} X_{\rm eff} }{\lambda^5} 
  - P_c - \delta P_{c,u}  \right )^2 \right ] \, , \\
\label{eq:BRSMKL} \cB( K_L \to \pi^0 \nu\bar\nu) = \kappa_L \left ( \frac{{\rm Im} 
    X_{\rm eff} }{\lambda^5} \right )^2 \, ,
\end{gather}
where
\be
X_{\rm eff} = V_{ts}^* V_{td} (X_{LL} + X_{RL}) 
\ee
and, as usual, 
$\lambda = |V_{us}|$, while $\kappa_+ = ( 5.173 \pm 0.025 ) \times 10^{-11} 
(\lambda/0.225)^8$~\cite{Mescia:2007kn} and 
$\kappa_L = ( 2.29 \pm 0.03 ) \times 10^{-10} (\lambda/0.225)^8$.
In the $K^+$ case the dimension-six charm quark corrections and subleading 
long-distance effects are characterized by 
$P_c = 0.372 \pm 0.015$~\cite{Brod:2008ss,Buras:2005gr,Buras:2006gb}
and $\delta P_{c,u} = 0.04 \pm 0.02$~\cite{Isidori:2005xm}, respectively.
The $X_{\rm eff}$ function can be rewritten as
\be
X_{\rm eff} = V_{ts}^* V_{td} X_{\rm SM} ( 1 +\xi e^{i\theta})
\ee
where $X_{\rm SM} =1.464 \pm 0.041$~\cite{Buras:2006gb} and
we have introduced the two real parameters $\xi$ and $\theta$ 
that vanish in the SM. The explicit expression
for these two parameters in our framework is 
\be
\xi e^{i\theta} =  - \frac{T}{ X_{\rm SM} }
\frac{ \Vt_{ts}^* \Vt_{td} }{ V_{ts}^* V_{td} }
\approx 5.6 \times 10^{2} \times \tc_{12} \ts_{12} e^{i(\phi_{21}^d+\beta-\beta_s)} c_{Z_R}^{\rm eff}~.
\label{eq:XRL2}
\ee

If we ignore the constraints from the $\Delta F=2$ processes 
there is certainly a large room for non-standard effects in $K \to \pi \nu\bar \nu$ decays, 
even taking into account the bound in Eq.~(\ref{eq:Zbs_lastb}). 
The situation changes if we implement the constraints on $\tc_{12}$ 
derived from $\varepsilon_K$, under the hypothesis of a large non-standard 
contribution to $S_{\psi\phi}$, analysed in Section~\ref{section:DFglobal}.
Here we should distinguish two cases:
\begin{figure}[t]
\begin{center}
\includegraphics[width=.6\textwidth]{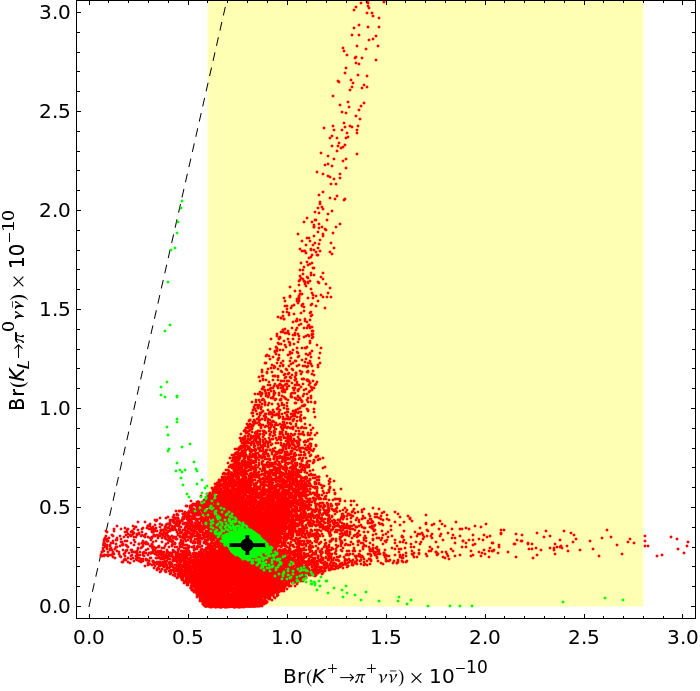}
\end{center}
\caption{\label{fig:Kpnn} 
Correlations between 
$\cB(K^+ \to \pi^+ \nu \bar \nu)$ and 
$\cB(K_L \to \pi^0 \nu \bar \nu)$ in our effective theory, taking into
account the $\varepsilon_K$ constraint. 
The red (dark) and green (light) points are obtained imposing 
the $\varepsilon_K$ constraint under the assumption of negligible 
or dominant contribution from $\mathcal{Q}_1^{LR}$, respectively.
The dashed line is the Grossman-Nir bound~\cite{Grossman:1997sk}. 
The vertical band correspond to the experimental result in~\cite{Artamonov:2008qb}
(1 $\sigma$ range) and the black cross to the SM prediction.}
\end{figure}
\begin{itemize}
\item{} If we can neglect the contribution 
of $\mathcal{Q}_1^{LR}$ to $\Delta F=2$ amplitudes, namely if we need 
to implement the bounds in Eqs.~(\ref{eq:c21fromRR}), then there is room 
for $\cO(1)$  deviations from the SM predictions in both rare modes for 
a sizable range of the phase $\phi^d_{12}$. This is shown by the red (dark) points
in Fig.~\ref{fig:Kpnn}, which are obtained imposing this constraint. 
As can be seen, larger deviations from the SM are also possible, but in this 
case the phase $\phi^d_{12}$ has to be tuned such that $\sin(2\phi^d_{12}) \approx 0$,
in order to avoid the  $\varepsilon_K$ constraint. This fined-tuned 
configuration gives rise to the two bands of red (dark) points in Fig.~\ref{fig:Kpnn}
with large enhancements of only one of the two branching ratios.
As noted in~\cite{Blanke:2009pq}
this structure is characteristic of all NP frameworks where the
phase in  $\Delta S=2$ amplitudes is the square of the CP-violating 
phase in $\Delta S=1$ FCNC amplitudes
(this is for instance what happens in the Little Higgs model 
with $T$ parity \cite{Blanke:2006eb}).

\item{}
If the contribution of $\mathcal{Q}_1^{LR}$ to $\Delta F=2$ amplitudes
is dominant, namely if we need to implement the stringent 
bounds in Eqs.~(\ref{eq:c21fromLR}), then the situation is more constrained. 
Here we can expect a visible deviation from the SM  only if the phase 
$\phi^d_{12}$ is tuned such that $\sin(\phi^d_{12}- \beta+\beta_s)\approx 0$,
where the bounds (\ref{eq:c21fromLR}) become less effective.
This give rise to the narrow band of green (light) points in  Fig.~\ref{fig:Kpnn}.
This correlation is very different from the one pointed out above, since in this
case the leading CP-violating phase in  $\Delta S=2$ amplitudes is not the 
square of the CP-violating phase appearing in $\Delta S=1$ FCNC amplitudes.
\end{itemize} 
If we relax the assumption of sizable NP contributions to $S_{\psi\phi}$ 
the predictions of these two modes do not change substantially: the
characteristic structures in  Fig.~\ref{fig:Kpnn} are indeed only due 
to the $\varepsilon_K$ constraint. As anticipated, the situation may change 
only if we could ignore the constraints from the $\Delta S=2$ processes,
assuming the corresponding Wilson coefficients are accidentally suppressed.
However, we consider this situation highly fine-tuned. Actually it should
be stressed that the maximal enhancements shown in  Fig.~\ref{fig:Kpnn}
also require a considerable amount of fine-tuning, both on the 
phase $\phi^d_{12}$ and on the ratio of $\Delta S=2$ over $\Delta S=1$
Wilson coefficients.

\subsection{Summary}
The main results of the $\Delta F=1$ analysis can be summarized as follows:
\begin{itemize}
\item{}
The constraints from $B_{s,d}\to\mu^+\mu^-$ eliminate the possibility 
of removing the known anomaly in the $Z\to b\bar b$ decay with the help
of right-handed currents.
\item{}
Contributions from RH currents to $B_{s,d}\to\mu^+\mu^-$,
$B \to \{X_s, K, K^*\} \nu\bar \nu$, and $K \to \pi \nu\bar \nu$ 
can all be significant, although this is not a general 
prediction of the model. If the deviations from the SM 
are sizable, the effects exhibit interesting patterns 
of correlations.
\item{}
In the $B_{s}\to\mu^+\mu^-$, case the branching ratio can receive
an  $\cO(1)$ enhancement over its SM expectation, but it cannot get close 
to its present experimental bound due to the  constraint from $B\to X_s l^+l^-$.
If the RH contribution to $S_{\psi\phi}$ is large, no significant 
enhancement is expected in $B_{d}\to\mu^+\mu^-$.
\item{}
If the RH contribution to $S_{\psi\phi}$ is large,
the pattern of possible enhancement/suppression 
in $B \to \{K, K^*\} \nu\bar \nu$ is  unambiguous,
as shown in Fig.~\ref{fig:BKnn}. 
\item{}
The pattern of possible enhancement/suppression 
in the to $K \to \pi\nu\bar\nu$ modes is largely 
independent of possible RH contributions to $S_{\psi\phi}$,
but it could help to disentangle the case where 
non-standard $\Delta F=2$ amplitudes are dominated 
by RR or RL operators.
\end{itemize}

\section{Comparison with MFV and explicit LR models} 

\subsection{RH currents vs.~MFV: general considerations}
The two effective theories are apparently very similar:
the low-energy particle content is the same and in both cases
we have a flavour group broken only by two Yukawas. 
However, the flavour groups are different:
$SU(3)_{Q_L} \times SU(3)_{U_R} \times SU(3)_{D_R}$
in the MFV case~\cite{D'Ambrosio:2002ex}
 vs.~$SU(3)_L \times SU(3)_R$ in the present
framework. The larger flavour symmetry of the MFV set-up
implies a much more constrained structure. In particular:
\begin{itemize}
\item{} If the normalization of the  two Yukawa couplings is
the same as in the SM, as expected with a single Higgs doublet, 
we are in the so-called CMFV regime~\cite{Buras:2000dm}.
In this case  deviations from the SM are small in all observables.
In particular, there is no hope to explain a large 
$S_{\psi\phi}$ asymmetry, contrary to what happens with RH
currents. 
\item{}
The expectation of a vanishingly small $S_{\psi\phi}$ in the MFV 
case remains true also with two Higgs doublets and large value of 
$\tan\beta$ if we assume that the Yukawa couplings are the only 
sources of $CP$ violation, as originally assumed in~\cite{D'Ambrosio:2002ex}.
In this case the only large deviation from the SM is a 
potentially large $B_s\to\mu^+\mu^-$ rate that could easily 
be just below the present experimental bound.
As we have seen, in the RH framework $B_s\to\mu^+\mu^-$ 
could be enhanced over its SM expectation, but the enhancement 
cannot be as large as in the MFV case at large $\tan\beta$
because of the  $B_s\to X_s \ell^+\ell^-$ 
constraint.
\end{itemize}

\subsection{Comparison with MFV with flavour-blind phases}
The phenomenology of MFV models can be quite different from the case 
discussed above if one relaxes the assumption that the Yukawa 
couplings are the only sources of $CP$ violation, or in the 
GMFV framework, as denoted in \cite{Kagan:2009bn}.
In particular, it has been recently shown
that in a two Higgs doublet model (2HDM) with
MFV, large $\tan\beta$, and flavour-blind CP-violating phases,
it is possible to generate a large $S_{\psi\phi}$ asymmetry 
and, as a consequence, automatically soften the anomalies in 
$S_{\psi K_S}$ and $\varepsilon_K$ in a correlated manner~\cite{Buras:2010mh}. 
In this set-up the NP contributions responsible for a 
large $S_{\psi\phi}$ are due to the exchange of heavy neutral scalars.
It is then interesting to compare this solution to the  $S_{\psi\phi}$
problem with the one considered here, where NP contributions 
originate presumably from new heavy gauge bosons.

Concentrating first on the high energy scales, at which new particles are 
integrated out let us emphasize that the relevant scales of right-handed 
currents are by roughly an order of magnitude larger than the allowed 
masses of neutral scalars. Also the Lorentz structure of the operators 
generated at the high scale is different in these models.
As seen in Eqs.~(\ref{eq:78}) and (\ref{eq:93ab}),
the presence of RH currents selects from the list in (\ref{eq:1.1})
non-vanishing initial conditions for
$\mathcal{Q}_1^{VRR}$ and $\mathcal{Q}_1^{LR}$,
while the leading $SU(2)_L \times U(1)_Y$ invariant 
operator induced by Higgs exchanges at the high scale is
$\mathcal{Q}_2^{LR}$.
As a consequence of this different operator structure, the NP contributions 
from RH currents are governed by the RG parameters $P_1^{VRR}=P_1^{VLL}$
and $P_1^{LR}$, while the ones in the 2HDM by $P_2^{LR}$. It is then interesting
 to observe that model independently $P_1^{LR}$ and $P_1^{LR}$ are roughly 
of the same magnitude but opposite sign. While this sign difference is not 
relevant in view of unknown signs of the coefficients involved in a
model with RH currents, it could play some role when a concrete 
model with RH currents is analyzed.

In the case of the 2HDM with MFV the pattern of NP contributions to 
$K^0-\bar K^0$, $B_d^0-\bar B_d^0$ and $B_s^0-\bar B_s^0$ mixings is 
governed by external quark masses implying naturally largest effects in 
the $B_s^0-\bar B_s^0$, followed by an order of magnitude $\cO(m_d/m_s)$ 
smaller effects in $B_d^0-\bar B_d^0$ mixing and negligible effects in 
$\varepsilon_K$. Still the impact of NP on the $B_d^0-\bar B_d^0$ mixing, 
uniquely following from the requirement of fitting a large $S_{\psi\phi}$,
softens automatically the $\varepsilon_K$ anomaly through the increase of 
the true value of $\sin 2\beta$  resulting from the fit to $S_{\psi K_S}$.
It should also be noted that the increase of $\sin 2\beta$ in the 2HDM
automatically favors larger values of $|V_{ub}|$ than those found in
$B\to \pi e\nu$, but the model does not offer an explanation why the 
exclusive and inclusive determinations give 
different values of~$|V_{ub}|$.

As we have shown in the previous sections, the pattern
of deviations from the SM is significantly different 
in the case of RH currents, although some points are 
common to both frameworks. In particular:
\begin{itemize}
\item
RH currents provide a natural explanation of the different values of $|V_{ub}|$
following from inclusive and exclusive decays, selecting the inclusive
determination as the one giving the true value of $|V_{ub}|$. This 
value is significantly higher than the corresponding value obtained from
the SM fits and, when combined with the tree level measurement of 
the angle $\gamma$, implies $\sin 2\beta= 0.77 \pm 0.05$.
\item 
Similarly to the MFV-2HDM case, 
the modified determination of $\sin 2\beta$ removes the $\varepsilon_K$ 
anomaly even in the absence of direct NP contributions to
this observable.
\item
In contrast to the MFV-2HDM case, the RH current contributions to 
$B_d^0-\bar B_d^0$ mixing are constrained to be negligible. As a result
we should expect $S_{\psi K_S}=\sin 2\beta$, implying a value of
$S_{\psi K_S}$ significantly larger than what determined from experiments.
\item
The model has sufficient number of parameters that it can naturally
generate a large value of $S_{\psi\phi}$ without any conflict with
other data. Contrary to the MFV-2HDM case, the large $S_{\psi\phi}$  does 
not imply $B_{s} \to \ell^+\ell^-$ close to the present experimental limit.
Moreover $B_{d} \to \ell^+\ell^-$ receives only small NP contribution 
 so that the MFV relation between the the branching ratios for 
$B_{s} \to \ell^+\ell^-$ and $B_{d} \to \ell^+\ell^-$ can be strongly violated.
\end{itemize}
We conclude then that  both models provide interesting solutions to the 
existing anomalies but:
\begin{itemize}
\item
the 2HDM with MFV cannot provide the 
explanation of different values of $|V_{ub}|$ 
following from inclusive and exclusive semi-leptonic decays;
\item
the model with RH currents provides this explanation but this feature
combined with tiny contributions to $B_d^0-\bar B_d^0$ mixing implies a
$\approx 2 \sigma$ 
discrepancy between the predicted value of $S_{\psi K_S}$ and its
experimental determination.
\end{itemize}

\subsection{Comparison with explicit left-right models.}

As anticipated in the introduction, there exist several analyses 
of flavour observables in explicit left-right models (see 
e.g.~\cite{Zhang:2007da,Maiezza:2010ic} and references therein).
In all these papers flavour observables are a key ingredient to 
determine the bounds on the masses of the massive RH gauge bosons.
Most of the existing analyses are focused on the minimal version 
of the model~\cite{Mohapatra:1974gc,Mohapatra:1974hk,Senjanovic:1975rk},
that is characterized by the gauge group 
$SU(2)_L \times SU(2)_R \times U(1)_{B-L}$, by a
discrete symmetry connecting the two $SU(2)$ groups, 
and by the minimal choice for the  Higgs sector necessary 
to achieve the two-step breaking 
$SU(2)_L \times SU(2)_R \times U(1)_{B-L} \to  SU(2)_L \times U(1)_{Y} \to U(1)_{\rm e.m.}$. 
A more general analysis where the latter
two hypotheses are relaxed can be found in~\cite{Langacker:1989xa}.

While some features of our effective theory approach can be 
applied also to these explicit models, there are a few important 
distinctive features of our analysis, related to the assumptions
behind the Yukawa  interactions. 
\begin{itemize}
\item{}
In our approach we make no assumptions about the existence 
of a L$\leftrightarrow$R discrete symmetry under which the Yukawa 
interactions must be invariant. This symmetry, that is 
usually enforced in the minimal models,
force the Yukawa matrices to be exactly (or approximately) LR
symmetric~\cite{Maiezza:2010ic}. As a result, $\Vt$ 
should have a hierarchical structure identical (or very similar) 
to the one of the CKM matrix. While this is a nice feature 
as far as protecting FCNCs, it would prevent us to solve 
the $V_{ub}$ problem, which was one of the main motivation 
of our analysis. As far as the structure of $\Vt$ and 
its impact in charged-currents are concerned, 
our analysis is thus more general than existing analysis in 
explicit LR models.
\item{}
While we do not impose any L$\leftrightarrow$R  discrete symmetry,
we assume a minimal structure for the breaking of the 
flavour group. We assume only two independent Yukawa couplings, 
with an extra protective symmetry forcing them to act 
only in the up- and down-type sector in the dimension-four 
operators, as shown in Eq.~(\ref{eq:Yd4}). This extra assumptions
have been imposed to have an efficient suppression of FCNCs in the 
left-handed sector, and to avoid scalar FCNCs at the tree level.
While these assumptions are quite reasonable and can easily 
be implemented in explicit LR models, they 
do not represent the most general possibility.
It is worth to stress that some of our phenomenological conclusions,
such as the absence of NP effects in $B_d$ mixing, after we require 
large NP effects in $B_s$ mixing, do depend on this assumption.
\end{itemize}

\section{Conclusions}

The possibility that at very short distance scales the nature  
is left-right symmetric appears to be intriguing.
As at low 
energy scales the parity is maximally broken and charged weak interactions 
exhibit left-handed structure, the right-handed weak currents, if present in 
nature, must be coupled to new heavy gauge bosons that are at least by one
order of magnitude heavier than the $W^\pm$ and $Z$ bosons of the SM.

While such heavy gauge bosons could be discovered at the LHC in the coming 
years, they can also manifest themselves in low energy processes. In our 
model
they are represented by new effective operators containing right-handed 
currents with their Wilson coefficients encoding the information about 
the fundamental theory, in particular the relevant couplings.

In the present paper we have analyzed
the impact of right-handed currents in both charged- and neutral-current 
flavour-violating processes  by means of an effective 
theory approach. To this end we have assumed
a left-right symmetric flavour group, commuting with 
an underlying $SU(2)_L \times SU(2)_R \times U(1)_{B-L}$ global symmetry, 
broken only by two Yukawa couplings. Having identified the leading six 
dimension operators in this model, we performed a rather detailed analysis of
those low energy observables, which could help to support or falsify the 
presence of this NP at  scales probed soon directly by the LHC.

The central role in our model is played by a new unitary matrix $\tilde V$
 that controls flavour-mixing in the right-handed sector. 
Using the data on the tree level charged current transitions $u\to d$,
 $u\to s$, $b\to u$ and $b\to c$ and the unitarity of $\tilde V$ 
we could
determine the structure of this matrix and demonstrate, following other 
authors, that the tension between inclusive and exclusive determinations
of $|V_{ub}|$ can be solved with the help of right-handed currents. 
The resulting true value of $|V_{ub}|$ turns 
out to be $(4.1 \pm 0.2 )\times 10^{-3}$, 
in the ball park of inclusive determinations.
The novel feature of our analysis as compared with~\cite{Crivellin:2009sd}
is the determination of the full right-handed matrix, 
and not only its selected elements, making use of unitarity.
We also find that while RH currents are very welcome to solve the 
``$|V_{ub}|$ problem'' they do not have a significant impact on the 
determination of $|V_{cb}|$ (as also pointed out in~\cite{Feger:2010qc}).

 Having determined the size and the flavour structure of 
right-handed currents that is consistent with the present data on tree 
level processes and which removes the ``$|V_{ub}|$-problem'', 
we have investigated how this NP would manifest itself in 
neutral current processes, including particle-antiparticle mixing, 
$Z\to b \bar b$, $B_{s,d}\to \mu^+\mu^-$,
 $B \to \{X_s,K, K^*\}\nu\bar \nu$ and $K\to \pi\nu\bar\nu$
decays. We have also addressed the possibility to explain a 
non-standard CP-violating phase 
in $B_s$ mixing in this context, and made the comparison with other 
predictive new-physics frameworks addressing the same problem. 

The main messages from this analysis are as follows:
\begin{itemize}
\item{}
The presence of RH currents in the model, in conjunction with the
already present SM left-handed currents generates in addition to the 
operators with $(V+A)\times(V+A)$ Dirac structure, also left-right 
operators $(V-A)\times(V+A)$. The contributions of the latter are 
known to be strongly enhanced at low energies through renormalization 
group effects and in the case of $\varepsilon_K$ and $\Delta M_K$ through 
chirally enhanced hadronic matrix elements of $(V-A)\times (V+A)$ operators. 
Consequently these observables put severe constraints on the model 
parameters ( as also known from various studies 
in explicit LR models~\cite{Zhang:2007da}).
\item{}
The desire to generate large CP-violating effects in $B_s$-mixing, 
hinted for by the enhanced value of $S_{\psi\phi}$ observed by the 
CDF and D0 collaborations, in conjunction with the $\varepsilon_K$-constraint,
implies additional constraints on the shape of $\tilde V$. In particular 
$\tc_{12}\ll 1$ and consequently $\tilde s_{12}\approx 1$. 
The pattern of deviations from the SM in this model is then as follows.
\item{}
The $S_{\psi\phi}$ and $\varepsilon_K$ anomalies can be understood.
\item{}
As a consequence of the large value of $\tilde s_{12}$, 
it should be possible  
to resolve the presence of RH currents also in $s\to u$ 
charged-current transitions. Here RH currents imply 
a  $\cO(10^{-3})$ deviation in 
the determination of $|V_{us}|$ from $K\to \pi\ell\nu$
and $K\to \ell\nu$ decays.
\item{}
The ``true value'' of $\sin 2\beta$ determined in our framework, 
namely the determination of the CKM phase $\beta$ on the basis
of the tree-level processes only, and in particular of $|V_{ub}|$,
is   $\sin 2\beta=0.77\pm 0.05$. This result is roughly $2\sigma$ 
larger than the measured value $S_{\psi K_S}=0.672\pm0.023$. 
This is 
a property of any explanation of the ``$|V_{ub}|$-problem'' by means 
of RH currents, unless the value of $|V_{ub}|$ from inclusive 
decays will turn out to be much lower than determined presently.
In general, such  discrepancy could be solved by a negative new 
CP-violating phase in $B_d^0-\bar B_d^0$ mixing. However, we 
have demonstrated that this is not possible in the present framework once the 
$\varepsilon_K$ constraint is imposed and large $S_{\psi\phi}$ is 
required. 
Thus we point out that simultaneous explanation of the  ``$|V_{ub}|$-problem'' 
and of $S_{\psi K_S}=0.672\pm0.023$ is problematic through RH
currents alone.
\item{}
The present constraints from $B_{s,d}\to\mu^+\mu^-$ eliminate the possibility 
of removing the known anomaly in the $Z\to b\bar b$ decay with the help
of right-handed currents.
On top of it, the constraint from $B\to X_s l^+l^-$ precludes 
$B_{s}\to \mu^+\mu^-$ to be close to its present experimental bound.
Moreover NP effects in $B_{d} \to \ell^+\ell^-$ are found generally 
smaller than in $B_{s} \to \ell^+\ell^-$.
\item{}
Contributions from RH currents to 
$B \to \{X_s,K, K^*\} \nu\bar \nu$ and
$K\to\pi\nu\bar\nu$ decays can still be significant. 
Most important, the deviations from the SM in these decays 
would exhibit a well-defined pattern of correlations.
\end{itemize}

We have compared this NP scenario with the general 
MFV framework and with more explicit NP models. 
Particularly interesting is the comparison with the 
2HDM with MFV, large $\tan\beta$, and flavour-blind CP-violating 
phases, where the $S_{\psi\phi}$ and  $\varepsilon_K$ anomalies 
can also be accommodated~\cite{Buras:2010mh}. 
What clearly distinguishes these two models at low-energies
is how they face the ``$|V_{ub}|$-problem"
(which can be solved only in the RH case) and the 
``$\sin2\beta$--$S_{\psi K}$ tension'' (which can be   
softened only in the 2HDM case). 
But also the future results on rare $B$ and $K$ decays listed 
above could in principle help to distinguish 
these two general NP frameworks.

 Restricting the discussion to these two NP frameworks, it appears 
 that a model with an extended scalar sector and right-handed currents
 could provide solutions to all the existing tensions in flavour
 physics simultaneously. This possibility can certainly be realized
 in explicit left-right symmetric models, where an extended Higgs
 sector  is also required to break the extended gauge symmetry.
 However, these
 extensions contain many free parameters and clear cut conclusions on
 the pattern of flavour violation cannot be as easily reached as it was
 possible in the simple framework considered here and in~\cite{Buras:2010mh}.

 We are looking forward to the upcoming experiments at the LHC, future
 $B$ factories, and rare $K$ decay experiments,
 that should be able to shed
 more light on the role of right-handed currents in flavour physics.

\subsubsection*{Acknowledgments}
We thank Wolfgang Altmannshofer and  Tillmann Heidsieck for useful discussions.
G.I.~ would like to thank Jernej Kamenik and Martin Gonzalez-Alonso
for useful discussions in the early stage of this work.
We all thank the Galileo Galilei Institute for Theoretical Physics 
for the hospitality and partial support during the completion of this work. 
This research was partially supported by the Cluster 
of Excellence `Origin and Structure 
of the Universe', by the Graduiertenkolleg GRK 1054 of DFG,
by the German `Bundesministerium f\"ur Bildung und Forschung' under 
contract 05H09WOE, and by the EU Marie Curie Research Training Network
contract MTRN-CT-2006-035482 ({\em Flavianet}).

\bibliographystyle{My}
\bibliography{bibliography}

\providecommand{\href}[2]{#2}\begingroup\raggedright\begin{thebibliography}{10}

\bibitem{Pati:1974vw}
J.~C. Pati and A.~Salam {\em Phys. Rev. Lett.} {\bf 32} (1974) 1083.

\bibitem{Mohapatra:1974gc}
R.~N. Mohapatra and J.~C. Pati {\em Phys. Rev.} {\bf D11} (1975) 2558.

\bibitem{Mohapatra:1974hk}
R.~N. Mohapatra and J.~C. Pati {\em Phys. Rev.} {\bf D11} (1975) 566--571.

\bibitem{Senjanovic:1975rk}
G.~Senjanovic and R.~N. Mohapatra {\em Phys. Rev.} {\bf D12} (1975) 1502.

\bibitem{Senjanovic:1978ev}
G.~Senjanovic {\em Nucl. Phys.} {\bf B153} (1979) 334.

\bibitem{Zhang:2007da}
Y.~Zhang, H.~An, X.~Ji, and R.~N. Mohapatra {\em Nucl. Phys.} {\bf B802} (2008)
  247--279, [\href{http://xxx.lanl.gov/abs/0712.4218}{{\tt arXiv:0712.4218}}].

\bibitem{Maiezza:2010ic}
A.~Maiezza, M.~Nemevsek, F.~Nesti, and G.~Senjanovic
  \href{http://xxx.lanl.gov/abs/1005.5160}{{\tt arXiv:1005.5160}}.

\bibitem{Csaki:2003zu}
C.~Csaki, C.~Grojean, L.~Pilo, and J.~Terning {\em Phys. Rev. Lett.} {\bf 92}
  (2004) 101802, [\href{http://xxx.lanl.gov/abs/hep-ph/0308038}{{\tt
  hep-ph/0308038}}].

\bibitem{Nomura:2003du}
Y.~Nomura {\em JHEP} {\bf 11} (2003) 050,
  [\href{http://xxx.lanl.gov/abs/hep-ph/0309189}{{\tt hep-ph/0309189}}].

\bibitem{Barbieri:2003pr}
R.~Barbieri, A.~Pomarol, and R.~Rattazzi {\em Phys. Lett.} {\bf B591} (2004)
  141--149, [\href{http://xxx.lanl.gov/abs/hep-ph/0310285}{{\tt
  hep-ph/0310285}}].

\bibitem{Georgi:2004iy}
H.~Georgi {\em Phys. Rev.} {\bf D71} (2005) 015016,
  [\href{http://xxx.lanl.gov/abs/hep-ph/0408067}{{\tt hep-ph/0408067}}].

\bibitem{Crivellin:2009sd}
A.~Crivellin {\em Phys. Rev.} {\bf D81} (2010) 031301,
  [\href{http://xxx.lanl.gov/abs/0907.2461}{{\tt arXiv:0907.2461}}].

\bibitem{Chen:2008se}
C.-H. Chen and S.-h. Nam {\em Phys. Lett.} {\bf B666} (2008) 462--466,
  [\href{http://xxx.lanl.gov/abs/0807.0896}{{\tt arXiv:0807.0896}}].

\bibitem{D'Ambrosio:2002ex}
G.~D'Ambrosio, G.~F. Giudice, G.~Isidori, and A.~Strumia {\em Nucl. Phys.} {\bf
  B645} (2002) 155--187, [\href{http://xxx.lanl.gov/abs/hep-ph/0207036}{{\tt
  hep-ph/0207036}}].

\bibitem{Feger:2010qc}
R.~Feger, V.~Klose, H.~Lacker, T.~Lueck, and T.~Mannel
  \href{http://xxx.lanl.gov/abs/1003.4022}{{\tt arXiv:1003.4022}}.

\bibitem{Isidori:2009ww}
G.~Isidori \href{http://xxx.lanl.gov/abs/0911.3219}{{\tt arXiv:0911.3219}}.

\bibitem{Feldmann:2006jk}
T.~Feldmann and T.~Mannel {\em JHEP} {\bf 02} (2007) 067,
  [\href{http://xxx.lanl.gov/abs/hep-ph/0611095}{{\tt hep-ph/0611095}}].

\bibitem{Amsler:2008zzb}
{ Particle Data Group} Collaboration, C.~Amsler {\em et.~al.} {\em Phys. Lett.}
  {\bf B667} (2008) 1.

\bibitem{Antonelli:2010yf}
M.~Antonelli {\em et.~al.} \href{http://xxx.lanl.gov/abs/1005.2323}{{\tt
  arXiv:1005.2323}}.

\bibitem{Bernard:2007cf}
V.~Bernard, M.~Oertel, E.~Passemar, and J.~Stern {\em JHEP} {\bf 01} (2008)
  015, [\href{http://xxx.lanl.gov/abs/0707.4194}{{\tt arXiv:0707.4194}}].

\bibitem{Antonelli:2009ws}
M.~Antonelli {\em et.~al.} \href{http://xxx.lanl.gov/abs/0907.5386}{{\tt
  arXiv:0907.5386}}.

\bibitem{Okamoto:2005zg}
M.~Okamoto {\em PoS} {\bf LAT2005} (2006) 013,
  [\href{http://xxx.lanl.gov/abs/hep-lat/0510113}{{\tt hep-lat/0510113}}].

\bibitem{Bernard:2008dn}
C.~Bernard {\em et.~al.} {\em Phys. Rev.} {\bf D79} (2009) 014506,
  [\href{http://xxx.lanl.gov/abs/0808.2519}{{\tt arXiv:0808.2519}}].

\bibitem{Gambino:2010bp}
P.~Gambino, T.~Mannel, and N.~Uraltsev
  \href{http://xxx.lanl.gov/abs/1004.2859}{{\tt arXiv:1004.2859}}.

\bibitem{Bona:2009cj}
{ UTfit} Collaboration, M.~Bona {\em et.~al.} {\em Phys. Lett.} {\bf B687}
  (2010) 61--69, [\href{http://xxx.lanl.gov/abs/0908.3470}{{\tt
  arXiv:0908.3470}}].

\bibitem{Laiho:2009eu}
J.~Laiho, E.~Lunghi, and R.~S. Van~de Water {\em Phys. Rev.} {\bf D81} (2010)
  034503, [\href{http://xxx.lanl.gov/abs/0910.2928}{{\tt arXiv:0910.2928}}].

\bibitem{Aaltonen:2007he}
{ CDF} Collaboration, T.~Aaltonen {\em et.~al.} {\em Phys. Rev. Lett.} {\bf
  100} (2008) 161802, [\href{http://xxx.lanl.gov/abs/0712.2397}{{\tt
  arXiv:0712.2397}}].

\bibitem{Abazov:2008fj}
{ D0} Collaboration, V.~M. Abazov {\em et.~al.} {\em Phys. Rev. Lett.} {\bf
  101} (2008) 241801, [\href{http://xxx.lanl.gov/abs/0802.2255}{{\tt
  arXiv:0802.2255}}].

\bibitem{Abazov:2010hv}
{ D0} Collaboration, V.~M. Abazov {\em et.~al.}
  \href{http://xxx.lanl.gov/abs/1005.2757}{{\tt arXiv:1005.2757}}.

\bibitem{Buras:2000dm}
A.~J. Buras, P.~Gambino, M.~Gorbahn, S.~Jager, and L.~Silvestrini {\em Phys.
  Lett.} {\bf B500} (2001) 161--167,
  [\href{http://xxx.lanl.gov/abs/hep-ph/0007085}{{\tt hep-ph/0007085}}].

\bibitem{Buras:2001ra}
A.~J. Buras, S.~Jager, and J.~Urban {\em Nucl. Phys.} {\bf B605} (2001)
  600--624, [\href{http://xxx.lanl.gov/abs/hep-ph/0102316}{{\tt
  hep-ph/0102316}}].

\bibitem{Babich:2006bh}
R.~Babich {\em et.~al.} {\em Phys. Rev.} {\bf D74} (2006) 073009,
  [\href{http://xxx.lanl.gov/abs/hep-lat/0605016}{{\tt hep-lat/0605016}}].

\bibitem{Buras:2008nn}
A.~J. Buras and D.~Guadagnoli {\em Phys. Rev.} {\bf D78} (2008) 033005,
  [\href{http://xxx.lanl.gov/abs/0805.3887}{{\tt arXiv:0805.3887}}].

\bibitem{Buras:2010pz}
A.~J. Buras, D.~Guadagnoli, and G.~Isidori
  \href{http://xxx.lanl.gov/abs/1002.3612}{{\tt arXiv:1002.3612}}.

\bibitem{Brod:2010mj}
J.~Brod and M.~Gorbahn \href{http://xxx.lanl.gov/abs/1007.0684}{{\tt
  arXiv:1007.0684}}.

\bibitem{Barberio:2008fa}
{ Heavy Flavor Averaging Group} Collaboration, E.~Barberio {\em et.~al.}
  \href{http://xxx.lanl.gov/abs/0808.1297}{{\tt arXiv:0808.1297}}.

\bibitem{Bona:2007vi}
{ UTfit} Collaboration, M.~Bona {\em et.~al.} {\em JHEP} {\bf 03} (2008) 049,
  [\href{http://xxx.lanl.gov/abs/0707.0636}{{\tt arXiv:0707.0636}}].

\bibitem{:2009ec}
{ Tevatron Electroweak Working Group} Collaboration
  \href{http://xxx.lanl.gov/abs/0903.2503}{{\tt arXiv:0903.2503}}.

\bibitem{Ligeti:2010ia}
Z.~Ligeti, M.~Papucci, G.~Perez, and J.~Zupan
  \href{http://xxx.lanl.gov/abs/1006.0432}{{\tt arXiv:1006.0432}}.

\bibitem{Lunghi:2008aa}
E.~Lunghi and A.~Soni {\em Phys. Lett.} {\bf B666} (2008) 162--165,
  [\href{http://xxx.lanl.gov/abs/0803.4340}{{\tt arXiv:0803.4340}}].

\bibitem{Buras:2010mh}
A.~J. Buras, M.~V. Carlucci, S.~Gori, and G.~Isidori
  \href{http://xxx.lanl.gov/abs/1005.5310}{{\tt arXiv:1005.5310}}.

\bibitem{Buchalla:1995vs}
G.~Buchalla, A.~J. Buras, and M.~E. Lautenbacher {\em Rev. Mod. Phys.} {\bf 68}
  (1996) 1125--1144, [\href{http://xxx.lanl.gov/abs/hep-ph/9512380}{{\tt
  hep-ph/9512380}}].

\bibitem{:2005ema}
{ LEP and SLD Electroweak Working Groups} Collaboration {\em Phys. Rept.} {\bf
  427} (2006) 257, [\href{http://xxx.lanl.gov/abs/hep-ex/0509008}{{\tt
  hep-ex/0509008}}].

\bibitem{Buchalla:2000sk}
G.~Buchalla, G.~Hiller, and G.~Isidori {\em Phys. Rev.} {\bf D63} (2000)
  014015, [\href{http://xxx.lanl.gov/abs/hep-ph/0006136}{{\tt
  hep-ph/0006136}}].

\bibitem{Altmannshofer:2009ma}
W.~Altmannshofer, A.~J. Buras, D.~M. Straub, and M.~Wick {\em JHEP} {\bf 04}
  (2009) 022, [\href{http://xxx.lanl.gov/abs/0902.0160}{{\tt
  arXiv:0902.0160}}].

\bibitem{Buras:2006gb}
A.~J. Buras, M.~Gorbahn, U.~Haisch, and U.~Nierste {\em JHEP} {\bf 11} (2006)
  002, [\href{http://xxx.lanl.gov/abs/hep-ph/0603079}{{\tt hep-ph/0603079}}].

\bibitem{Buras:2003td}
A.~J. Buras {\em Phys. Lett.} {\bf B566} (2003) 115--119,
  [\href{http://xxx.lanl.gov/abs/hep-ph/0303060}{{\tt hep-ph/0303060}}].

\bibitem{Aaltonen:2007kv}
{ CDF} Collaboration, T.~Aaltonen {\em et.~al.} {\em Phys. Rev. Lett.} {\bf
  100} (2008) 101802, [\href{http://xxx.lanl.gov/abs/0712.1708}{{\tt
  arXiv:0712.1708}}].

\bibitem{Abazov:2007iy}
{ D0} Collaboration, V.~M. Abazov {\em et.~al.} {\em Phys. Rev.} {\bf D76}
  (2007) 092001, [\href{http://xxx.lanl.gov/abs/0707.3997}{{\tt
  arXiv:0707.3997}}].

\bibitem{Kamenik:2009kc}
J.~F. Kamenik and C.~Smith {\em Phys. Lett.} {\bf B680} (2009) 471--475,
  [\href{http://xxx.lanl.gov/abs/0908.1174}{{\tt arXiv:0908.1174}}].

\bibitem{Bartsch:2009qp}
M.~Bartsch, M.~Beylich, G.~Buchalla, and D.~N. Gao {\em JHEP} {\bf 11} (2009)
  011, [\href{http://xxx.lanl.gov/abs/0909.1512}{{\tt arXiv:0909.1512}}].

\bibitem{Barate:2000rc}
{ ALEPH} Collaboration, R.~Barate {\em et.~al.} {\em Eur. Phys. J.} {\bf C19}
  (2001) 213--227, [\href{http://xxx.lanl.gov/abs/hep-ex/0010022}{{\tt
  hep-ex/0010022}}].

\bibitem{:2007zk}
{ BELLE} Collaboration, K.~F. Chen {\em et.~al.} {\em Phys. Rev. Lett.} {\bf
  99} (2007) 221802, [\href{http://xxx.lanl.gov/abs/0707.0138}{{\tt
  arXiv:0707.0138}}].

\bibitem{:2008fr}
{ BABAR} Collaboration, B.~Aubert {\em et.~al.} {\em Phys. Rev.} {\bf D78}
  (2008) 072007, [\href{http://xxx.lanl.gov/abs/0808.1338}{{\tt
  arXiv:0808.1338}}].

\bibitem{Altmannshofer:2008dz}
W.~Altmannshofer {\em et.~al.} {\em JHEP} {\bf 01} (2009) 019,
  [\href{http://xxx.lanl.gov/abs/0811.1214}{{\tt arXiv:0811.1214}}].

\bibitem{Buras:2005gr}
A.~J. Buras, M.~Gorbahn, U.~Haisch, and U.~Nierste {\em Phys. Rev. Lett.} {\bf
  95} (2005) 261805, [\href{http://xxx.lanl.gov/abs/hep-ph/0508165}{{\tt
  hep-ph/0508165}}].

\bibitem{Isidori:2005xm}
G.~Isidori, F.~Mescia, and C.~Smith {\em Nucl. Phys.} {\bf B718} (2005)
  319--338, [\href{http://xxx.lanl.gov/abs/hep-ph/0503107}{{\tt
  hep-ph/0503107}}].

\bibitem{Mescia:2007kn}
F.~Mescia and C.~Smith {\em Phys. Rev.} {\bf D76} (2007) 034017,
  [\href{http://xxx.lanl.gov/abs/0705.2025}{{\tt arXiv:0705.2025}}].

\bibitem{Brod:2008ss}
J.~Brod and M.~Gorbahn {\em Phys. Rev.} {\bf D78} (2008) 034006,
  [\href{http://xxx.lanl.gov/abs/0805.4119}{{\tt arXiv:0805.4119}}].

\bibitem{Grossman:1997sk}
Y.~Grossman and Y.~Nir {\em Phys. Lett.} {\bf B398} (1997) 163--168,
  [\href{http://xxx.lanl.gov/abs/hep-ph/9701313}{{\tt hep-ph/9701313}}].

\bibitem{Artamonov:2008qb}
{ E949} Collaboration, A.~V. Artamonov {\em et.~al.} {\em Phys. Rev. Lett.}
  {\bf 101} (2008) 191802, [\href{http://xxx.lanl.gov/abs/0808.2459}{{\tt
  arXiv:0808.2459}}].

\bibitem{Blanke:2009pq}
M.~Blanke {\em Acta Phys. Polon.} {\bf B41} (2010) 127,
  [\href{http://xxx.lanl.gov/abs/0904.2528}{{\tt arXiv:0904.2528}}].

\bibitem{Blanke:2006eb}
M.~Blanke {\em et.~al.} {\em JHEP} {\bf 01} (2007) 066,
  [\href{http://xxx.lanl.gov/abs/hep-ph/0610298}{{\tt hep-ph/0610298}}].

\bibitem{Kagan:2009bn}
A.~L. Kagan, G.~Perez, T.~Volansky, and J.~Zupan {\em Phys. Rev.} {\bf D80}
  (2009) 076002, [\href{http://xxx.lanl.gov/abs/0903.1794}{{\tt
  arXiv:0903.1794}}].

\bibitem{Langacker:1989xa}
P.~Langacker and S.~Uma~Sankar {\em Phys. Rev.} {\bf D40} (1989) 1569--1585.

\end{thebibliography}\endgroup
\end{document}